\documentstyle[multicol,prd,psfig,aps]{revtex}

\begin{document}
\draft
\preprint{TIFR/TH/00-04, IMSc/2000/01/02}
\title{A QCD Analysis of Polarised Parton Densities}
\bigskip
\author{Dilip Kumar Ghosh and Sourendu Gupta}
\address{Department of Theoretical Physics,\\ Tata Institute of Fundamental Research,\\
    Homi Bhabha Road, Bombay 400005, India.}
\author{D.\ Indumathi}
\address{The Institute of Mathematical Sciences,\\ CIT Campus, Chennai 600113, India.}
\maketitle
\begin{abstract}
We present the results of QCD fits to global data on deep-inelastic
polarised lepton-hadron scattering.  We find that it is possible to fit
the data with strongly broken $SU(2)$ flavour for the polarised sea
densities.  This can be tested in $W$ production at polarised
RHIC.  The data fails to pin down polarised singlet sea quark and gluon
densities. We explore the uncertainties in detail and show that
improvement in statistics, achievable at polarised HERA for
measurement of $A_1$ at moderately low values of $x$, have large payoffs
in terms of the improvement in measurement of gluon densities.
\end{abstract}
\section{Introduction}

It is now more than a decade since the first polarised DIS experiments
\cite{emc} discovered the strong breaking of a $SU(6)$ quark model based
sum-rule \cite{ejsr}, and precipitated the ``proton spin crisis''.  Since
then many polarised deep-inelastic scattering (DIS) experiments have
reported measurements of the virtual photon asymmetry
\begin{equation}
   A_1(x,Q^2) = \frac{g_1(x,Q^2)}{F_1(x,Q^2)}
\label{defa1}\end{equation}
on different targets (the structure functions $g_1$ and $F_1$ are defined
later).  A polarised proton collider at RHIC will soon begin to constrain
the unknown polarised parton distributions even more strongly. Current
and future interest in this topic stems partly from the history of the
``spin crisis''.

However, polarised parton densities are also interesting because of
the role they might play in future polarised hadron collider searches
for completions of the standard model. Essentially, a large variety of
physics beyond the standard model plays with chirality. Some of this
freedom can easily be curtailed by polarised scattering experiments,
if the polarised parton densities are known with precision. We expect
that by the end of the polarised-RHIC program this goal should be
reached.

The longitudinally polarised structure function $g_1$ is defined by
\begin{equation}
   g_1(x,Q^2) = {\widetilde {\cal C}}_q\otimes \frac12\sum_f e_f^2
      \left[{\widetilde q}_f + {\widetilde{\bar q}}_f\right] +
    \frac12\left(\sum_f e_f^2\right)
            {\widetilde {\cal C}}_g\otimes {\widetilde g},
\label{defg1}\end{equation}
which is a Mellin convolution of the quark (${\widetilde q}_f$),
anti-quark (${\widetilde{\bar q}}_f$) and  gluon ($\widetilde g$)
longitudinally polarised distributions with the coefficient functions
${\widetilde {\cal C}}_{q,g}$. The index $f$ denotes flavour, and $e_f$
is the charge carried by the quark. The unpolarised structure function
is given by a similar formula in terms of the corresponding unpolarised
densities and coefficient functions.

These coefficient functions and the splitting functions (which determine
the evolution of the densities) are computable order by order in
perturbative QCD.  The former are crucial for the Bjorken sum rule
\cite{bjsr}, connecting the proton and neutron structure functions,
$g_1^p$ and $g_1^n$, to the neutron $\beta$-decay constant $g_A$, and
are known to NNLO \cite{nnlo}.  This makes it possible to use this sum
rule for precision measurement of the QCD scale \cite{altarelli}. The
polarised splitting functions are known only at NLO \cite{pol-nlo}.

In this paper we analyse the currently available inclusive DIS data
in QCD and extract polarised parton distributions from them.  In this
respect our work is similar to that of \cite{prevfit}. However, our
analysis differs in several ways.  For one, some of the data we use is
more recent than the older fits. More importantly, we relax some of the
assumptions which needed to be made in analysing the older data. We allow
for flavour asymmetry in the polarised sea quark densities, and let the
first moment of the gluon density vary freely in the fit. Furthermore
we make a detailed investigation of the uncertainties in these polarised
gluon and sea quark densities.

The uncertainty in gluon densities may seem puzzling in view of the fact
that the $Q^2$-dependence of the structure function $g_1$ involves the
gluon strongly. In fact, at LO we already have--- \begin{equation}
   \frac{\partial g_1}{\partial\log Q^2} = \frac{\alpha_S}{2\pi}
     \left[{\widetilde P}_{qq}\otimes g_1 +
           {\widetilde P}_{qg}\otimes{\widetilde g}\right],
\end{equation}
where ${\widetilde P}_{qq}$ and ${\widetilde P}_{qg}$ are polarised
splitting functions.  Since $\alpha_S$ is now very strongly constrained
by measurements at LEP and through unpolarised DIS, one might expect that
data on $g_1$ constrains the polarised gluon densities. Unfortunately,
errors on $g_1$ are large in the low-$x$ region, where the contribution
of the gluons dominate, primarily because the asymmetry $A_1$ is small
at low-$x$. For the same reason the flavour singlet sea quark density is
also rather loosely constrained by data.  We investigate the statistics
necessary to improve these constraints through DIS measurements of $A_1$
at polarised HERA.

The plan of this paper is the following. In the next section we discuss the
various technicalities that distinguish different global analyses. This
section also serves to set up the notation. This is followed by a section
that discusses our choice of data used in the fit. The next section contains
our results for the LO and NLO fits, and a detailed consideration of the
parameter errors. A section on some applications of our parametrisations
follows this. The final section contains a summary of our main results.

\section{Constraints on parton densities}

\subsection{Parton densities and structure functions}

With $N_f$ flavours of quarks, we need to fix $2N_f+1$ parton densities.
These are for the $2N_f$ flavours of quarks and anti-quarks and the gluon.
For protons or neutrons, the quark and anti-quark densities for the
strange and heavier flavours are equal.  We work with the two (flavour
non-singlet) polarised valence quark densities, ${\widetilde V}_u$ and
${\widetilde V}_d$, corresponding to the up and down flavours. The other
non-singlet densities we use are those corresponding to the diagonal
generators of $SU(5)$ flavour---
\begin{eqnarray}
\nonumber
   {\widetilde q}_3\equiv 2(\widetilde{\bar u}-\widetilde{\bar d}),
        \qquad&&\qquad
   {\widetilde q}_8\equiv 2(\widetilde{\bar u}+\widetilde{\bar d}
                           -2\widetilde{\bar s}),\\
   {\widetilde q}_{15}\equiv 2(\widetilde{\bar u}+\widetilde{\bar d}
                               +\widetilde{\bar s}-3\widetilde{\bar c}), 
        \qquad&&\qquad
   {\widetilde q}_{24}\equiv 2(\widetilde{\bar u}+\widetilde{\bar d}
                               +\widetilde{\bar s}+\widetilde{\bar c}
                              -4\widetilde{\bar b}).
\label{densities}\end{eqnarray}
The initial conditions for evolution are that below and at each flavour
threshold, the density for that flavour of quarks is zero. Thus,
below the charm threshold we have ${\widetilde q}_{24}={\widetilde
q}_{15}={\widetilde q}_0$ and below the bottom threshold we set
${\widetilde q}_{24}={\widetilde q}_0$. For the singlet quark density,
we use
\begin{equation}
   {\widetilde q}_0\equiv2\sum_f{\widetilde{\bar q}}_f=
                  \widetilde\Sigma-\sum_f{\widetilde V}_f,
\label{singlet}\end{equation}
in preference to the usual $\widetilde\Sigma$ (which is the sum over
quark and anti-quark densities of all flavours). The evolution
equations couple ${\widetilde q}_0$ to the gluon density $\widetilde
g$. We also define similar unpolarised quark and gluon
densities\footnote{Our convention is that polarised quantities are
distinguished from the corresponding unpolarised ones by a tilde.}.

Finally, the structure functions $g_1^p$ and $g_1^n$, for the proton
and neutron, are given by eq.\ (\ref{defg1}). The unpolarised structure
functions $F_1^{p,n}$ are given by the analogous expression where the
polarised parton densities are replaced by the unpolarised densities.
An isospin flip, interchanging ${\widetilde V}_u$ and ${\widetilde
V}_d$ and switching the sign of ${\widetilde q}_3$, relates $g_1^p$
and $g_1^n$. After correcting for nuclear effects, the normalised
structure function for deuterium is $g^d_1=(g^p_1+g^n_1)/2$, and a
similar expression for $F^d_1$.

We shall have occasion to use the first moments of various polarised
parton distributions. We introduce the notation---
\begin{eqnarray}
\nonumber
    \Gamma_u(Q^2) = \int_0^1 dx\,{\widetilde V}_u(x,Q^2),
       \qquad&&\qquad
    \Gamma_d(Q^2) = \int_0^1 dx\,{\widetilde V}_d(x,Q^2), \\
    \Gamma_i(Q^2) = \int_0^1 dx\,{\widetilde q}_i(x,Q^2),
       \qquad&&\qquad
    \Gamma_g(Q^2) = \int_0^1 dx\,{\widetilde g}(x,Q^2). \\
\end{eqnarray}
We will also use the notation $\Gamma_V^0=\Gamma_u+\Gamma_d$ and
$\Gamma_V^3=\Gamma_u-\Gamma_d$. The notation
\begin{equation}
   \Gamma_1^{n,p}(Q^2) = \int_0^1 dx\, g_1^{n,p}(x,Q^2)
\end{equation}
is fairly standard. We shall use it in the text. We also use the
notation $\Gamma_{\bar u}$, {\sl etc.\/}, to denote the first
moments of the flavoured sea densities.

\subsection{The fitting strategy}

We use experimental data on the asymmetries $A_1^p$, $A_1^n$ and $A_1^d$,
measured on proton, neutron (${}^3{\rm He}$) and deuterium targets to constrain
the polarised parton densities. We assume full knowledge of the unpolarised
parton densities as given by some global fit, so that the structure
function $F_1$ can be reconstructed using appropriate NLO coefficient
functions. Then the data on $A_1$ can be
converted to $g_1$. We prefer this method to taking the $g_1$ values
presented by experiments, since different experimental groups may make
different assumptions about the unpolarised structure functions. Such
effects would lead to additional normalisation uncertainties in any
global fit.

We have chosen to use the CTEQ4 set of parton densities \cite{cteq4} in
our work.  We do not expect this choice to affect our conclusions strongly
since the unpolarised parton densities now have smaller errors than data
on the polarisation asymmetry $A_1$. However, with this choice we are
constrained to follow some of the assumptions made by the CTEQ group---
\begin{enumerate}
\item
   We work in the $\overline{\rm MS}$ scheme, since the CTEQ group does that.
   Other possibilities would have been to work in the AB \cite{ab} or JET
   \cite{jet} schemes, but then we would have had to transform the CTEQ
   distributions. We prefer to avoid this procedure, since the best fit parton
   densities in one scheme do not necessarily transform into the best fit
   densities in another scheme.
\item
   We retain the CTEQ choice for the charm quark mass being 1.6 GeV and the
   bottom quark mass to be 5.0 GeV. At each mass threshold, we increase the
   number of flavours by one, and treat the newly activated flavour as
   massless immediately above the threshold. Parton distributions and
   $\alpha_S$ are continuous across these thresholds \cite{matching}.
\item
   We are constrained to use the $\Lambda_{QCD}$ values used in \cite{cteq4}.
\item
   We take $Q_0^2=2.56$ GeV${}^2$ in order to avoid having to evolve the
   unpolarised parton densities downwards.
\end{enumerate}

\noindent
In future we plan to study the results of relaxing one or more of these
restrictions.

We follow the parametrisation of CTEQ4 and write---
\begin{equation}
   {\widetilde f}(x,Q_0^2) = a_0 x^{a_1} (1-x)^{a_2} (1+a_3 x^{a_4}),
\label{param}\end{equation}
for all densities apart from $q_3$, which is parametrised as
\begin{equation}
   {\widetilde q}_3(x,Q_0^2) = a_0 x^{a_1} (1-x)^{a_2} (1+a_3 \sqrt x + a_4 x).
\label{param3}\end{equation}
We have made the choice that the large-$x$ behaviour of any polarised
density is the same as that of the unpolarised density; in other words,
the parameter $a_2$ is the same for corresponding polarised and unpolarised
densities (this assumption is sometimes given the name ``helicity
retention property'' \cite{largex}).  For simplicity we have also equated
the polarised and unpolarised values of $a_4$ when this parameter is a power.

Finally, at $Q_0^2$ we have extended some of the CTEQ assumptions for
unpolarised parton densities to polarised. These include equating
the values of $a_1$ for ${\widetilde V}_u$, ${\widetilde V}_d$ and
${\widetilde q}_3$, taking $a_4=1$ for ${\widetilde q}_0$, equating the
values of $a_2$ for ${\widetilde q}_0$ and ${\widetilde q}_3$. We also
retain the choice $\eta\equiv2{\widetilde s}/(\widetilde{\bar u}+
\widetilde{\bar d})=1/2$ in some of our fits, but let it vary in others.
Although these assumptions seem overly
restrictive, the quality of the data does not allow us to fit many of
these parameters. We discuss some of these points later in this paper.

The main difference between our parametrisation and previous ones is
that we explicitly include a non-zero ${\widetilde q}_3(x,Q_0^2)$ and
break $SU(2)$ flavour symmetry in the polarised sea. This part of the
sea density is actually quite well constrained, and plays a crucial role
in our fits.

\subsection{Sum rules}

In a three flavour world, we can write down the following sum rule for
the first moments of the nucleon structure functions in NLO QCD---
\begin{equation}
   \Gamma_1^{p,n} = \left(\pm\frac16g_3+\frac1{18}g_8+\frac19g_0\right)
          \left\{ 1 - \frac{\alpha_S}{\pi} \right\}.
\label{nloej}\end{equation}
The anomalous dimensions on the right hand side of this equation
have been calculated in the $\overline{\rm MS}$ scheme \cite{nnlo}.
Two facts used are that in the $\overline{\rm MS}$ scheme the first
moment of the NLO quark coefficient ${\widetilde {\cal C}}_q^{(1)}=-2$,
and the first moment of the gluon coefficient function is ${\widetilde
{\cal C}}_g^{(1)}=0$.  The upper and lower signs belong to protons
and neutrons, respectively.  The quantities $g_3$, $g_8$ and $g_0$
are baryonic axial couplings.  They are defined as matrix elements of
axial vector currents between baryon states.  Due to the axial anomaly,
the singlet axial-vector current is not conserved.  As a result, $g_0$
picks up a $Q^2$ dependence \cite{jaffe}.  Hence, $g_0$, the moments, and
$\alpha_S$ have to be evaluated at the same $Q^2$ in eq.\ (\ref{nloej}).

It is not easy to extract $g_0$ from low-energy hadron data, although
there have been some attempts to do this using elastic $\nu p$ scattering
\cite{g0}. This gives $g_0=0.14\pm0.27$.  Lattice computations \cite{g0l}
and QCD sum rules \cite{g0s} also give similar numbers, but have
systematic uncertainties which have to be removed in future.  The quantity
$g_3=1.2670\pm0.0035$ \cite{pdg} is obtained from the neutron beta-decay
constant.  The coupling $g_8$ is extracted from the decay of strange to
non-strange baryons.  $SU(3)$ flavour symmetry is used crucially in this
extraction \cite{g8}. The PDG result is $g_8=0.579\pm0.025$ \cite{pdg}.

Using the coefficient functions in the $\overline{\rm MS}$ scheme,
the moments of the structure functions can be expressed as
\begin{equation}
   \Gamma_1^{p,n}(Q^2) = \left\{1 - \frac{\alpha_S}{\pi}\right\} \left[
      \frac5{18}\Gamma_V^0(Q^2)\pm\frac16\biggl(\Gamma_V^3(Q^2)+\Gamma_3(Q^2)\biggr)
                           +\frac1{18}\Gamma_8(Q^2)+\frac29\Gamma_0(Q^2)\right],
\label{nloap}\end{equation}
where the upper (lower) sign is for the proton (neutron). Apart from eq.\
(\ref{defg1}), we have used the definitions of the non-singlet densities
in eq.\ (\ref{densities}) and the singlet parton density in eq.\
(\ref{singlet}).

There are two sum rules which can be obtained by equating the right-hand
sides of eqs.\ (\ref{nloej}) and (\ref{nloap}). Alternatively, we could
use some linear combinations. The only one that removes the coupling $g_0$
is the difference $\Gamma_1^p-\Gamma_1^n$, and
gives the Bjorken sum rule \cite{bjsr}.
At NLO this is---
\begin{equation}
    \Gamma_V^3 + \Gamma_3 = g_3.
\label{nlobj}\end{equation}
Note that the first moments of non-singlet densities are independent of
$Q^2$ order by order to all orders. As a result, this equation is also
valid in this form to all orders.  We impose this form of the Bjorken
sum rule on our fits.

The sum of the two gives the Ellis-Jaffe sum rule when the additional
assumption $g_8=g_0$ is made. This cannot be correct in QCD because $g_0$
is $Q^2$ dependent and $g_8$ is not. Moreover, in the absence of a real
measurement of $g_0(Q^2)$, an Ellis-Jaffe type sum rule cannot constrain
the parton densities as the Bjorken sum rule does.  Hence, we use such
a sum rule to extract $g_0$ rather than to impose it as a constraint on
parton densities.

\subsection{Positivity}

Polarisation asymmetries are the ratios of the difference and sum
of physically measurable cross sections. Since cross sections
are non-negative, asymmetries are bounded by unity in absolute value.
In the parton model or in LO QCD, these cross sections are directly
related to parton densities. Hence positivity of cross sections imply
\begin{equation}
   \left|\frac{{\widetilde f}(x,Q^2)}{f(x,Q^2)}\right| \le 1
\label{positive}\end{equation}
for the ratio of each polarised and unpolarised density to leading
order in QCD. In our LO fits, we impose these restrictions.

At NLO and beyond, this simple relation between parton densities and cross
sections no longer holds. Parton densities are renormalisation scheme
dependent objects; although universal, they are not physical. Hence they
need not satisfy positivity \cite{positive}, instead one must impose
positivity on the actual cross sections. This is numerically difficult
and requires knowledge of a variety of cross sections evaluated to NLO.
Since this knowledge is lacking, and for numerical simplicity, we have
instead imposed eq.\ (\ref{positive}) on all our parton density fits.

\subsection{Choice of numerical techniques}

Our numerical goal is to evolve parton density functions with absolute
errors of at most $10^{-3}$. If this design goal were reached, then
numerical errors would lie at least an order of magnitude below all
other errors.  We integrate the evolution equations using a 4-th order
Runge-Kutta algorithm. The Mellin convolutions required in the evaluation
of the derivative are computed using a Gauss-Legendre integral. The
parton densities are evaluated on a grid and interpolated using a cubic
spline method. All the numerical algorithms may be found in \cite{numrec}.

The knot points of the cubic spline are selected to give an accuracy
of $10^{-5}$ in the evaluation of the parton densities.  The Mellin
convolutions are also accurate to this order. We require the Runge-Kutta
to give us integration errors bounded by $10^{-4}$.  This gives us the
error limits we require. We can test these estimates by checking that
all sum rules are satisfied to within $10^{-3}$. On a 180 MHz R10000
processor, the program takes about 0.15 CPU seconds to evolve the parton
densities by $\Delta Q^2=1$ GeV${}^2$.

\section{Selection of data}

Experiments do not measure the asymmetry $A_1$ directly; they measure
the asymmetry between the cross sections for lepton and longitudinally
polarised hadrons being parallel and anti-parallel---
\begin{equation}
   A_L = \frac{d\sigma(\uparrow\uparrow)-d\sigma(\uparrow\downarrow)}
              {d\sigma(\uparrow\uparrow)+d\sigma(\uparrow\downarrow)},
\end{equation}
or a similar asymmetry, $A_T$, with transversely polarised hadrons.
These asymmetries are related to the two that we require by
\begin{equation}
   A_L=D(A_1+\eta A_2) \qquad{\rm and}\qquad A_T=d(A_2-\xi A_1),
\end{equation}
where $D$ and $d$ are depolarisation factors for the virtual photon
and $\xi$ and $\eta$ are essentially kinematic constants. In terms of
the ratio of the Compton scattering cross sections for transversely and
longitudinally polarised virtual photons,
\begin{equation}
   R=\frac{\sigma_{\scriptscriptstyle L}}{\sigma_{\scriptscriptstyle T}}
    = \frac{F_2-2xF_1}{2xF_1},
\end{equation}
we can write
\begin{equation}
   D = \frac{y(2-y)}{y^2+2(1-y)(1+R)}, \qquad{\rm and}\qquad
   \eta = 2\gamma\frac{1-y}{2-y}.
\end{equation}
Here $\gamma=2Mx/Q\ll1$. Using the degree of transverse polarisation
of the virtual photon,
\begin{equation}
   \epsilon = \frac{1-y}{1-y+y^2/2},
\end{equation}
we can write
\begin{equation}
   d= D\sqrt{\frac{2\epsilon}{1+\epsilon}}, \qquad{\rm and}\qquad
   \xi = \eta\frac{1+\epsilon}{2\epsilon}.
\end{equation}
Since $\gamma$ is very small in the DIS region, the relations $A_L=DA_1$ and
$A_T=dA_2$ are actually satisfied to high accuracy.
We then use the further relations
\begin{equation}
   A_1 = (g_1-\gamma^2 g_2)/F_1 \qquad{\rm and}\qquad
   A_2 = \gamma(g_1 + g_2)/F_1,
\label{a12}\end{equation}
to obtain eq.\ (\ref{defa1}) when $\gamma\ll1$. It is clear from the
second equation that $g_2$ is difficult to measure.

The main theoretical uncertainty in measurements of $A_1$ is in the values
of $R$ used. In fact, many experiments use $R$ in two ways. First, it
enters the expression for $D$ and $d$, and hence is used to compute
QED corrections for initial state radiation\footnote{We thank Abhay
Deshpande for drawing our attention to this point.} and to construct
$A_1$ and $A_2$ from $A_L$ and $A_T$. Next, it is used along with
measurements of $F_2$ to compute $F_1$ and thus relate $g_1$ to $A_1$.
We bypass this second use of $R$ by utilising experimental data on $A_1$
instead of $g_1$.  We are forced, however, to accept the first use of $R$,
In any case, differences between experiments in their estimates of $D$
should be factored into the overall normalisation errors.

The SMC collaboration has data from muon scattering off both proton and
deuterium targets. Data was taken in separate runs in 1993 and 1996. The
most recent publication for $A_1$ is \cite{smc}; this supersedes
previously published data.  The E-143 experiment at SLAC has data from
electron scattering off proton, deuterium and ${}^3{\rm He}$ targets. Their most
recent publication is \cite{e143}, which supersedes all previous published
data on $A_1(x,Q^2)$ by this collaboration.  The E-154 experiment at
SLAC has data from electron scattering off ${}^3{\rm He}$ targets \cite{e154}.
The HERMES collaboration in DESY has data from positron scattering
off protons and ${}^3{\rm He}$ \cite{hermes}. We have also used data on
DIS from ${}^3{\rm He}$ taken by the SLAC E-142 collaboration \cite{e142}.
We have chosen not to utilise data taken by the older EMC collaboration
and the E-140 experiments at SLAC.

Deuterium is a spin-1 nucleus with the $p$ and $n$ primarily in a relative
$s$-wave state. The $d$-wave probability is estimated to be $\omega_D=
0.05\pm0.01$ \cite{deuteron}. This is used in the relation between the
structure function of deuterium and those of $p$ and $n$---
$g_1^d=(1-3\omega_D/2)(g_1^p+g_1^n)/2$. In ${}^3{\rm He}$, the two protons are
essentially paired into a spin singlet, and the asymmetry is largely due
to the unpaired neutron. Corrections due to other components of the
nuclear wave-function are small \cite{helium}. More details are available
in \cite{review}.

From the chosen experiments, we have retained only the data on
$A_1(x,Q^2)$ for $Q^2\ge2.56$ GeV${}^2$. While this does remove some of
the low-$x$ data, the error bars in the removed data are pretty large. We
have checked by backward evolution that the data which is removed would
not have constrained the fits any further. The total number of data
points used in our analysis is 224.

In most fitting procedures the statistical errors on measurements are
combined in some way with the systematic error estimates. Both sets of
errors are usually reported in the literature in each
bin of data. Whereas this procedure is acceptable for statistical
errors, it oversimplifies the nature of systematic errors. These latter
are correlated from bin to bin, and one must use the full covariance
matrix of errors in the analysis. In the absence of published information
on the covariance matrix, one may make the simplifying assumption that
the bin-to-bin correlation vanishes, and add the statistical and
systematic errors in quadrature. This overestimates the errors on data
and hence the errors on the parameters determined by fitting. We have
made a different extremal assumption of neglecting the systematic errors
altogether. This procedure almost certainly leads us to under-estimate the
parameter errors--- a point to be borne in mind when we discuss large
errors and uncertainties in the fits. In summary, our choice of error
analysis is deliberately conservative.

Since $g_2$ contains a possible twist-3 contribution, which cannot be
written in terms of parton distributions, we cannot utilise data on
$g_2$ for our fits. However, the twist-2 part is completely determined
by $g_1$.  In a later section, we report an attempt to limit the extent
of the twist-3 term using our fitted polarised parton densities. For
this we have utilised data on proton target from SMC and the E-143
collaboration at SLAC \cite{g2p}, on deuterium target from SMC,
E-143 and SLAC E-155 \cite{g2d}, and on neutron target from the E-143
and SLAC experiment E-154 \cite{g2n}.  In all cases, we have used the
most recent data set and analysis from each collaboration.  The quality
of data on $g_2$ is poorer than that for $g_1$. This is because $A_2$
is small, and extraction of $g_2$ from $A_2$ requires the subtraction
of $g_1$, which itself has significant measurement errors.  The errors
are dominated by statistical uncertainties.

There remains data from semi-inclusive DIS taken by the SMC \cite{semi-s}
and HERMES \cite{semi-h} experiments. Analysis of these require
fragmentation functions and their $Q^2$ evolution. Since such analyses
are still to reach the stage that parton densities have, utilising
parametrisations for fragmentation functions would introduce larger
errors into our fits. For this reason, we have chosen not to use such
data in this work.

\section{Results}

We have made four full analyses--- two LO and two NLO, each with and
without $SU(2)$ flavour symmetry (denoted S and $\overline{\rm S}$
respectively) for the sea quarks, and with fixed $\eta=0.5$. In addition,
we have made a set of fits with $SU(2)$ symmetric sea ($\widetilde q_3=0$)
but $\eta$ allowed to vary freely. At LO this had no effect on the fit---
$\eta=0.5$ gave the best fit and the remaining parameters were identical
to LO S. At NLO $\eta$ moved to 0.6, but the parameters remained close to
the set NLO S.  The goodness of fit improved only marginally when $\eta$
was allowed to float. Given the uncertainties in the remaining parameters
we retain the choice $\eta=0.5$ in the main work reported below.

\begin{figure}[hbtp]
\begin{center} \leavevmode
   \psfig{figure=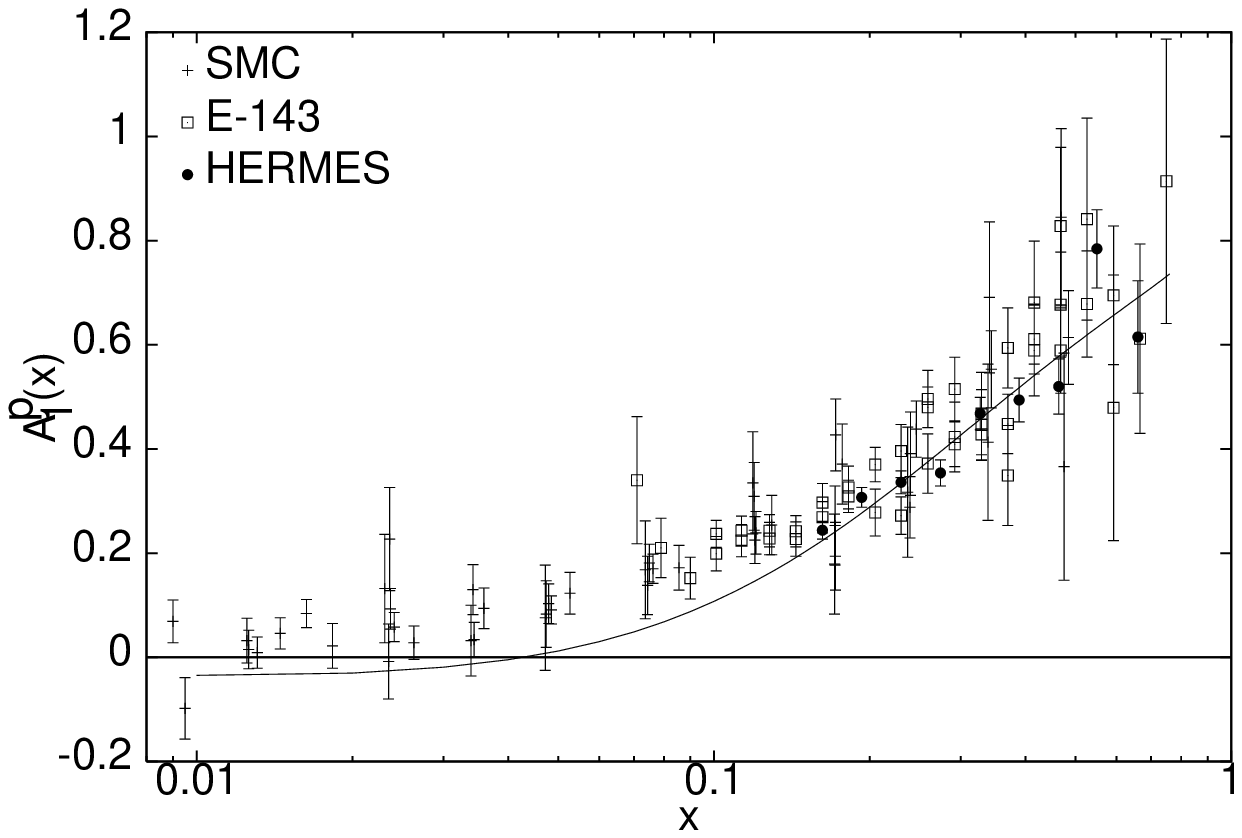,height=5.5cm,width=7.9cm}
\end{center}
\caption[dummy]{Data and fits for $A_1^p$. The data are at different $Q^2$, but
    the curve is the asymmetry from NLO $\overline{\rm S}$ set calculated at
    fixed $Q_0^2=5$ GeV${}^2$.}
\label{fg.a1p}\end{figure}

\begin{figure}[hbtp]
\begin{center} \leavevmode
   \psfig{figure=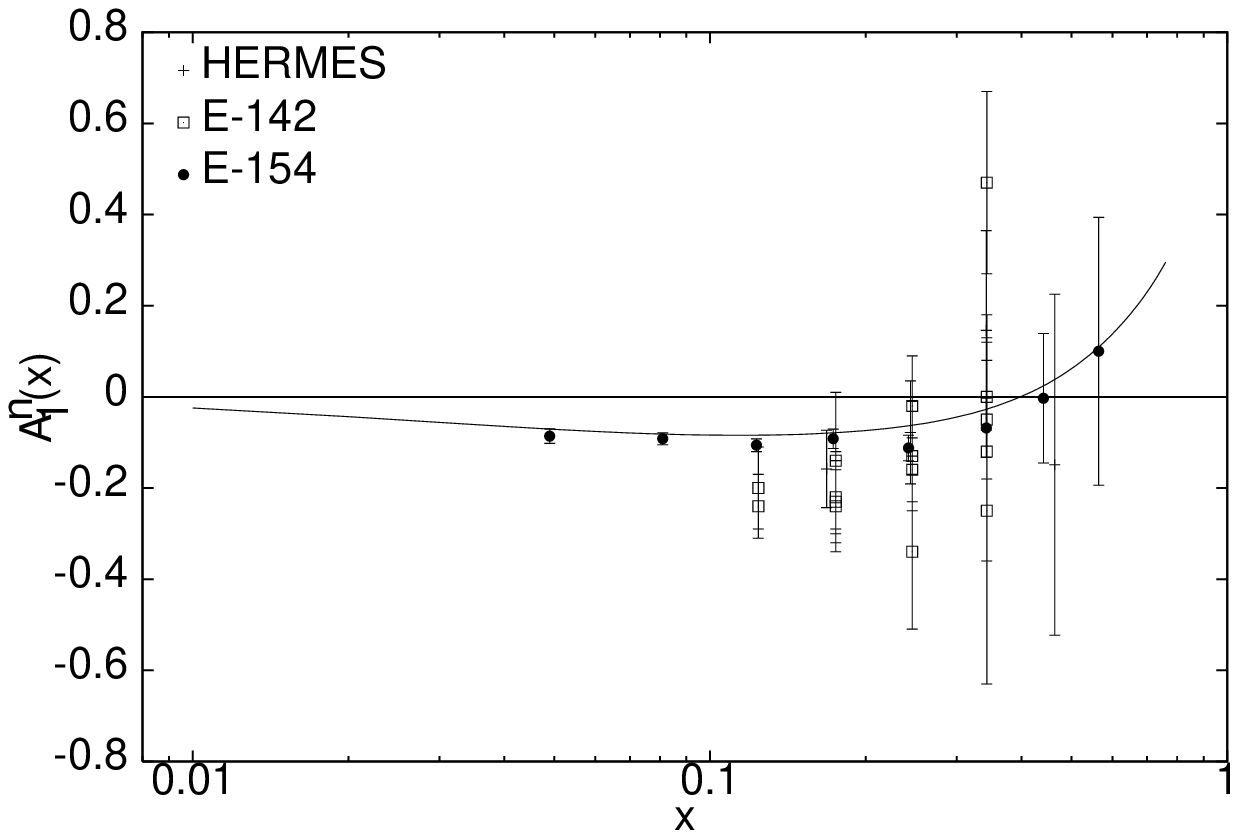,height=5.5cm,width=7.9cm}
\end{center}
\caption[dummy]{Data and fits for $A_1^n$. The data are at different $Q^2$, but
    the curve is the asymmetry from NLO $\overline{\rm S}$ set calculated at
    fixed $Q_0^2=5$ GeV${}^2$.}
\label{fg.a1n}\end{figure}

The goodness of fit, and the constraints imposed by each set of data are
summarised in Table \ref{tb.chi2}. The values of $\chi^2$ favour the
NLO sets slightly. The HERMES proton and the E-154 neutron data (see
Figures \ref{fg.a1p}, \ref{fg.a1n}) express the strongest preferences
for the NLO fits; almost the entire change in $\chi^2$ in going from
LO to NLO comes from these two data sets. It is probably no coincidence
that these two data sets also have the smallest error bars.

\begin{figure}[hbtp]
\begin{center} \leavevmode
   \psfig{figure=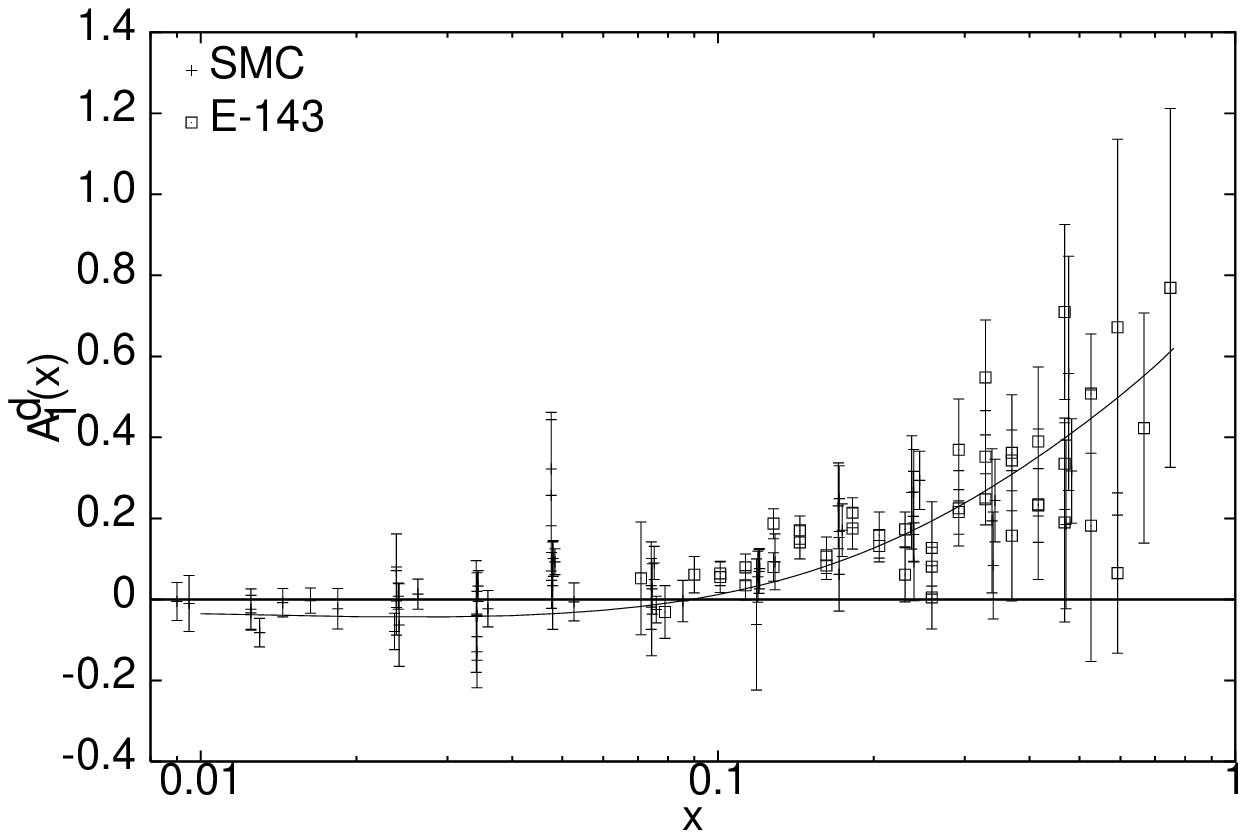,height=5.5cm,width=7.9cm}
\end{center}
\caption[dummy]{Data and fits for $A_1^d$. The data are at different $Q^2$, but
    the curve is the asymmetry from NLO $\overline{\rm S}$ set calculated at
    fixed $Q_0^2=5$ GeV${}^2$.}
\label{fg.a1d}\end{figure}

The parameters and their error estimates are shown in Tables
\ref{tb.nlo}--\ref{tb.lo0}. It is worth noting that the NLO $\overline{\rm
S}$ densities lie well within the limits allowed by eq.\ (\ref{positive}),
as do all the densities for NLO S except the gluon. The parameters $a_0$
and $a_3$ for $\widetilde g$ in NLO S are at the limit of positivity.
The quality of the NLO $\overline{\rm S}$ fit is shown in Figures
\ref{fg.a1p}--\ref{fg.a1d}.  We recommend that the NLO parametrisations be
used with the CTEQ4M set of unpolarised parton densities and the LO with
the CTEQ4L set \cite{cteq4}, and with appropriate values of $\Lambda_{QCD}$.

For the $\overline{\rm S}$ densities, the normalisation of ${\widetilde
q}_3$ inherits its error from the valence parameters and the coupling
$g_3$. It is the best constrained among the parameters describing the
sea. Similarly, for the two S densities the normalisation of ${\widetilde
V}_d$ is fixed by the Bjorken sum rule and its error is inherited from
the remaining valence parameters.

\begin{figure}[hbtp]
\begin{center} \leavevmode
   \psfig{figure=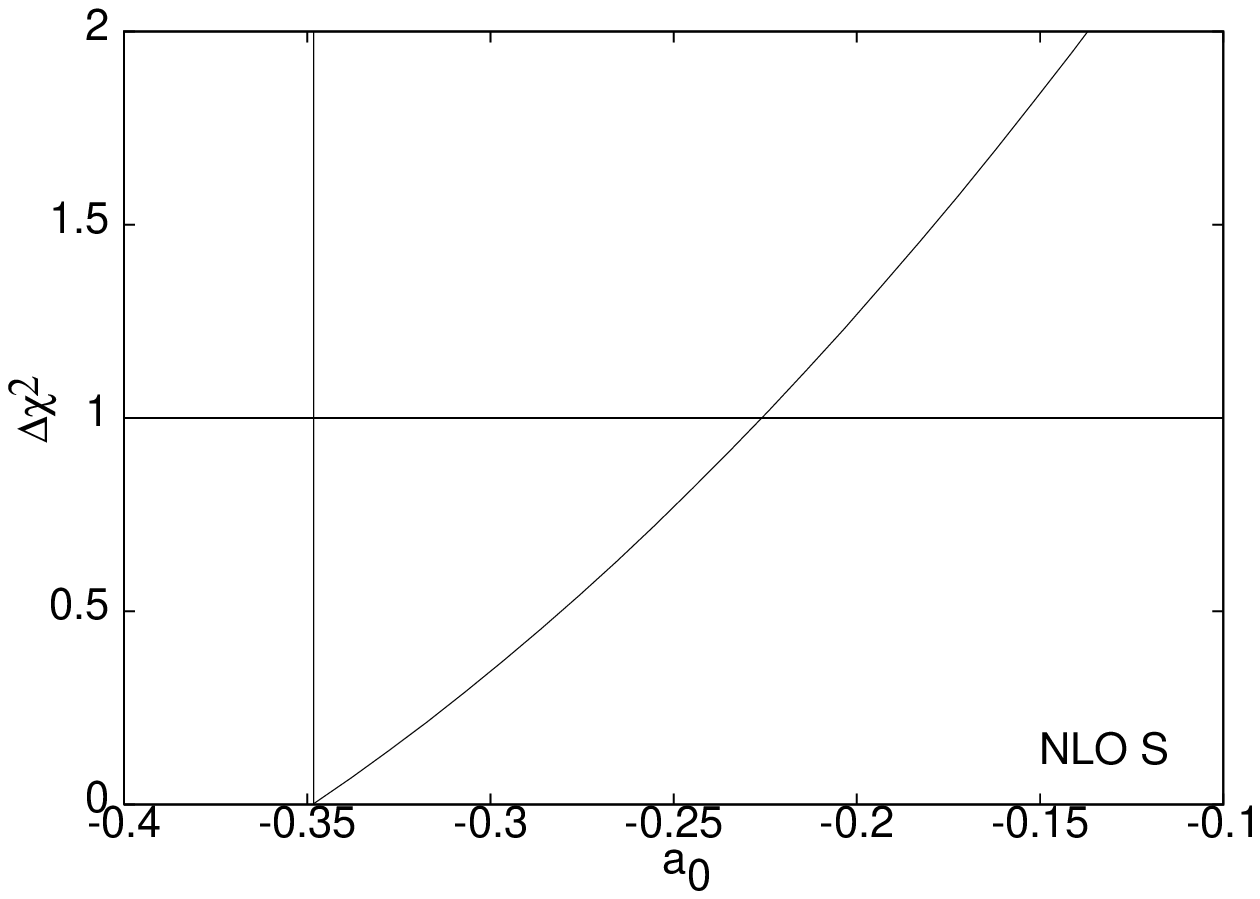,height=5cm,width=7cm}
   \psfig{figure=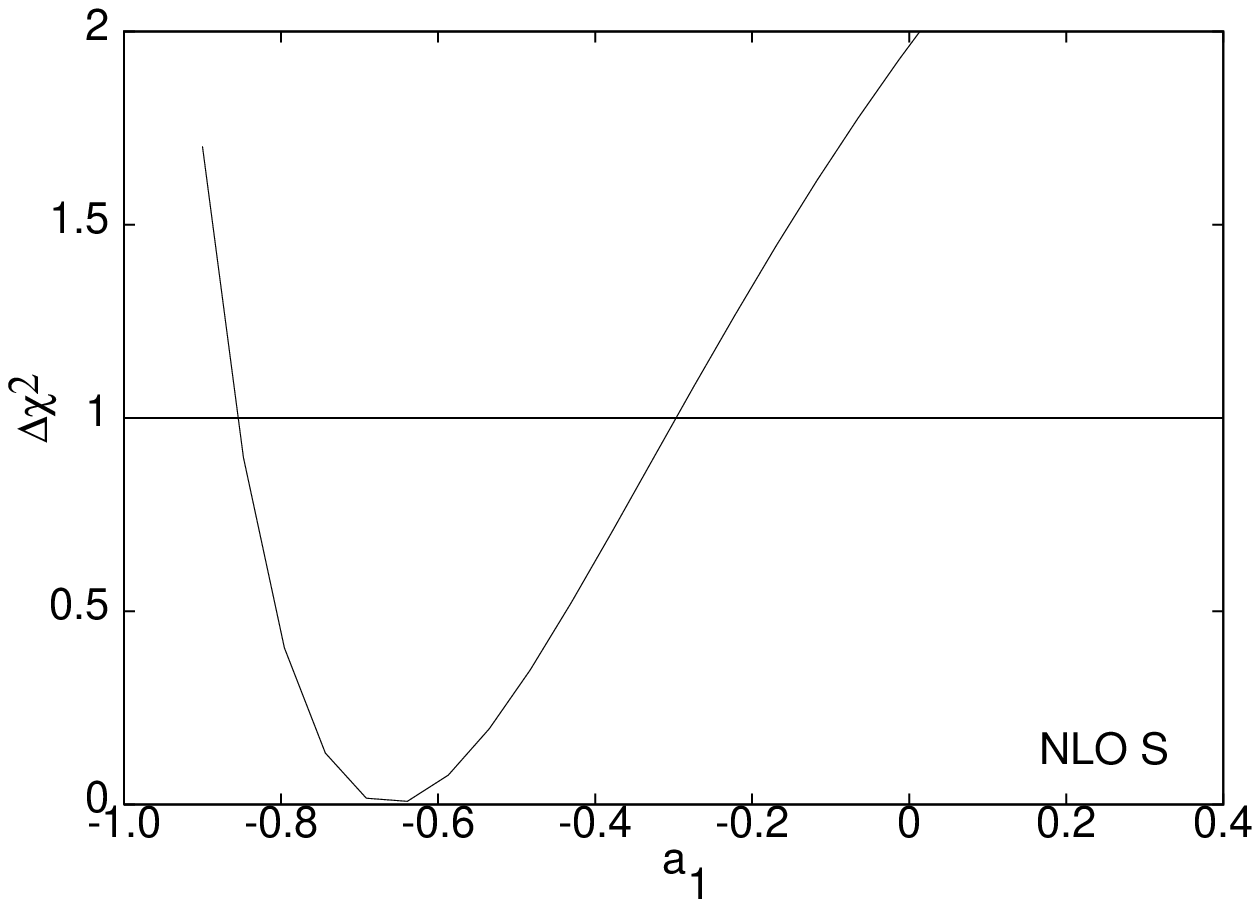,height=5cm,width=7cm}
   \psfig{figure=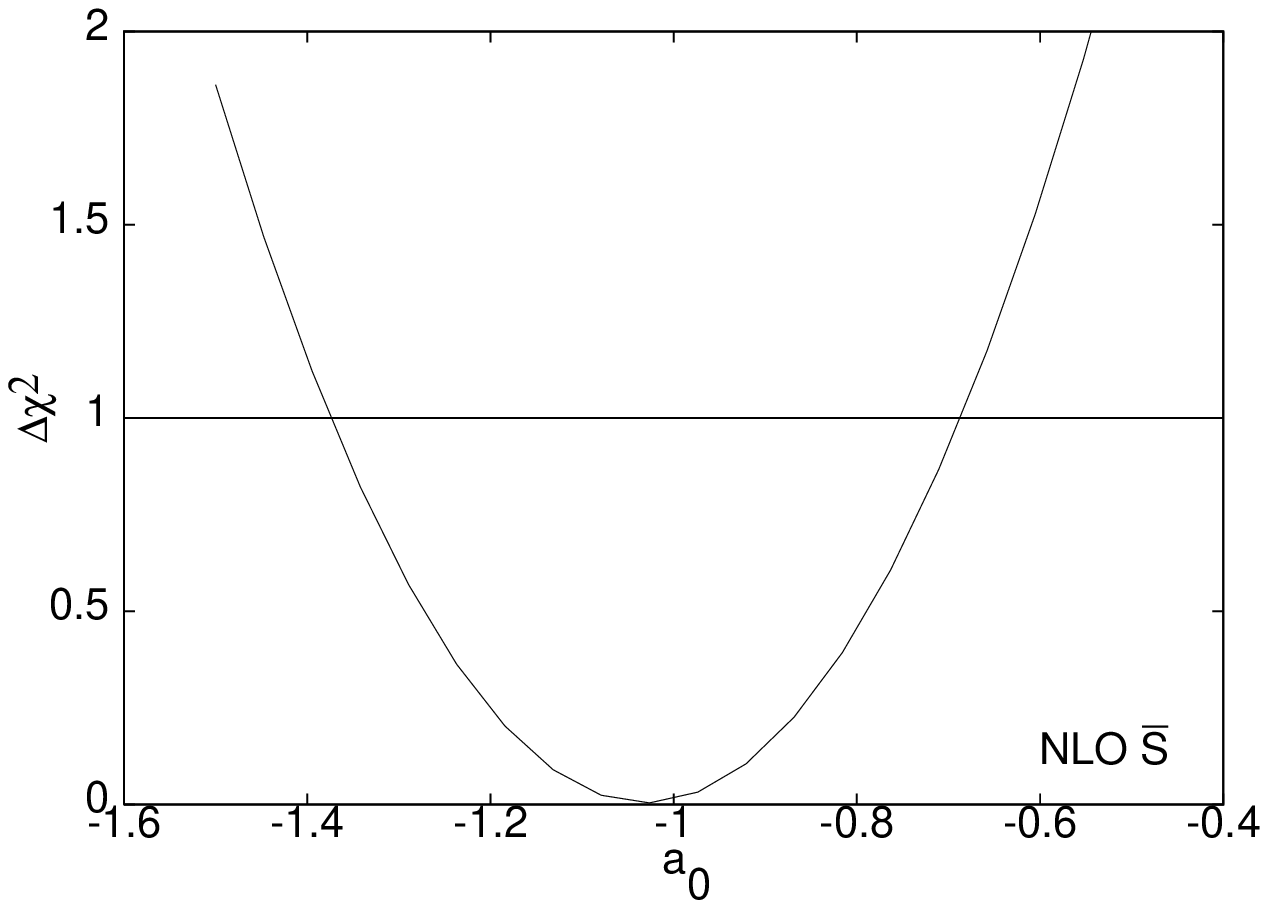,height=5cm,width=7cm}
   \psfig{figure=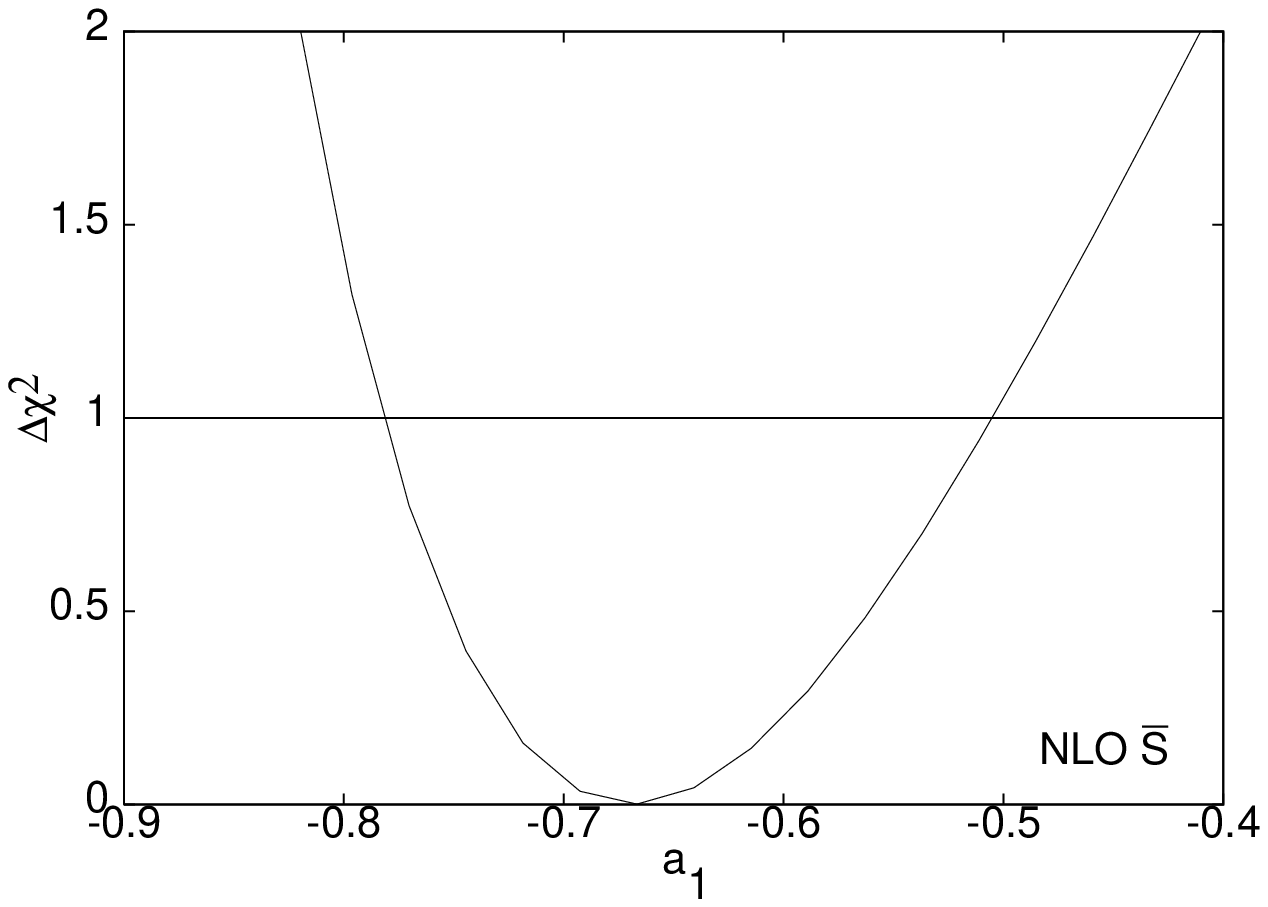,height=5cm,width=7cm}
\end{center}
\caption[dummy]{Plots of $\Delta\chi^2$ against the parameters $a_0$ and $a_1$
    for $\widetilde g$ in the two NLO sets. In the set NLO S, $a_0$ is at the
    boundary of positivity.}
\label{fg.gln}\end{figure}

\begin{figure}[hbtp]
\begin{center} \leavevmode
   \psfig{figure=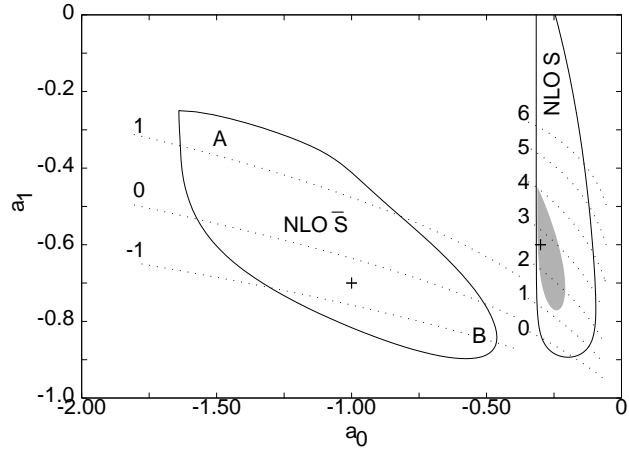,height=6cm,width=8.5cm}
\end{center}
\caption[dummy]{The covariance of the fitted parameters $a_0$ and $a_1$
    for $\widetilde g$ for the two NLO sets with $a_3$ kept at their
    respective best fit values. The crosses show the best fit
    points, and full lines are the contours enclosing the 68\% confidence
    limits. The reference points A and B are used to quantify the variation
    in gluon densities in Figure \protect{\ref{fg.parton}}. Along the dashed
    lines $\Gamma_g=-1/2^n$, for the values of $n$ marked. The grey patch is
    the region allowed by the faked data discussed later.}
\label{fg.glncr}\end{figure}

In Fig \ref{fg.gln} we have shown the variation of
\begin{equation}
   \Delta\chi^2 = \chi^2 - \chi^2_{\rm min}
\end{equation}
when one of the parameters in $\widetilde g$ is varied for fixed values
of all the other parameters in the set. The minimum of these curves fixes
the best-fit value of the parameter, and the points where $\Delta\chi^2=1$
give the 68.3\% confidence limits on this parameter.  Uncertainties in the
gluon density are shown in greater detail in Figure \ref{fg.glncr}. This
shows contour lines of $\Delta\chi^2=2.3$, which encloses the area with
68.3\% probability of giving a good description of the data. Also shown in
the figure are lines of constant $\Gamma_g(Q_0^2)$. Although $1\sigma$
contours give negative values of $\Gamma_g$, we note that $3\sigma$
contours include positive values as well. Note that our error bars are
deliberately conservative, due to our neglect of systematic errors.
Since systematic errors are as large as the statistical errors, the
naive procedure of summing them in quadrature would have led us to
believe that at NLO positive $\Gamma_g$ is allowed at $1.5\sigma$.

\begin{figure}[hbtp]
\begin{center} \leavevmode
   \psfig{figure=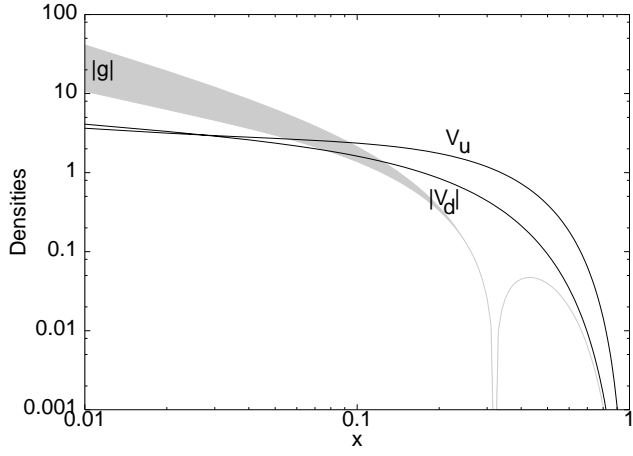,height=6cm,width=8.5cm}
\end{center}
\caption[dummy]{The absolute values of the polarised valence quark densities
    are shown along with the range of allowed $|\widetilde g|$ in the NLO
    $\overline{\rm S}$ set. The gray band showing this uncertainty is the
    band enclosed by the densities obtained at points A and B marked in
    Figure \protect{\ref{fg.glncr}}.}
\label{fg.parton}\end{figure}

The huge uncertainties in the gluon distribution due to these parameter
variations are illustrated in Figure \ref{fg.parton}.  At $x=0.01$ the
polarised gluon density at NLO can lie anywhere in the range from $-10$
to $-50$. This uncertainty at low-$x$ prevents us from investigating this
theoretically interesting region. The situation is worse at LO where
$\widetilde g$ is at the limit of integrability, since $a_1\approx-1$.
Thus $\Gamma_g$ is essentially undetermined.

\begin{figure}[hbtp]
\begin{center} \leavevmode
   \psfig{figure=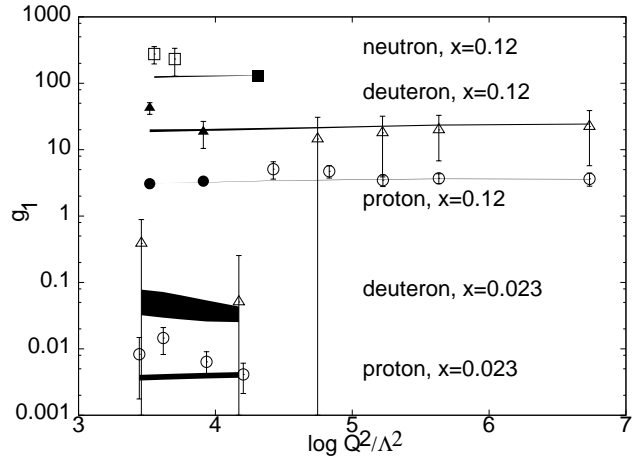,height=6cm,width=8.5cm}
\end{center}
\caption[dummy]{We show $g_1(x,Q^2)$ in selected bins of $x$ as a function
    of $\log(Q^2/\Lambda^2)$. The data on $g_1^p$ from SMC are shown by open circles
    and from E-143 by filled circles; on $g_1^d$ from the same experiments by open
    and filled triangles respectively, and on $g_1^d$ from E-142 by open squares
    and from E-154 by filled squares. The bands are the uncertainty in $g_1$ induced
    by the uncertainty in $\widetilde g$ shown in Figure \protect{\ref{fg.parton}}.
    Data in different bins of $x$ are offset vertically for visibility.}
\label{fg.qsq}\end{figure}

This large uncertainty in $\widetilde g$ comes because the
only constraint on gluon densities at present are the data on $Q^2$
variations of $g_1$. Furthermore, the data at $x<0.1$ are most effective
in constraining $\widetilde g$, and in this range, the data have large
errors. We have quantified this in Figure \ref{fg.qsq}, where we show
selected data on $g_1$ in different bins of $x$ as a function of $Q^2$
(the normalisation has been made arbitrary for ease of viewing).  As the
bands of variation due to $\widetilde g$ (in the NLO $\overline{\rm S}$
set) show, the data does not constrain the gluon density well.

A few qualitative statements are in order. It is clear that in a
scaling theory, DIS data with nucleon or nuclear targets can at most
fix two linear combinations of quark densities. However, DIS structure
functions are $Q^2$-dependent. Hence, in principle, data of arbitrarily
high accuracy fixes these two linear combinations at each $Q^2$---
{\sl i.e.\/}, an infinite number of functions.  In QCD there is a more
economical description of this $Q^2$ dependence involving the set of
parton densities given in Section 2. Of course, in the real world data
are never infinitely precise, so the question is how accurately does
$Q^2$ evolution fix these densities. Some part of the answer is clear
from Figure \ref{fg.qsq}--- improved data at low $x$ will constrain
$\widetilde g$ much better and $g_1^d$, being iso-singlet, would present
the best constraint.

\begin{figure}[hbtp]
\begin{center} \leavevmode
   \psfig{figure=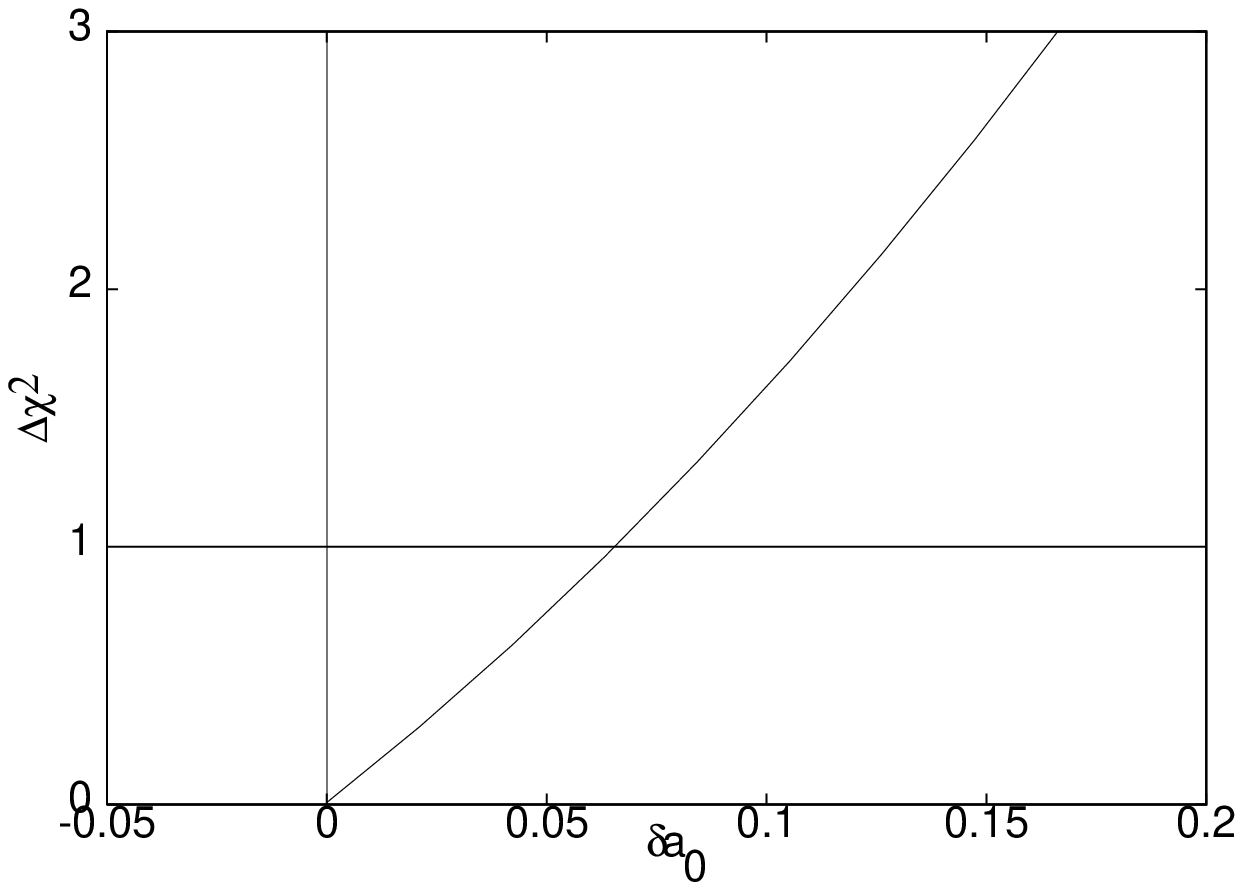,height=5cm,width=7cm}
   \psfig{figure=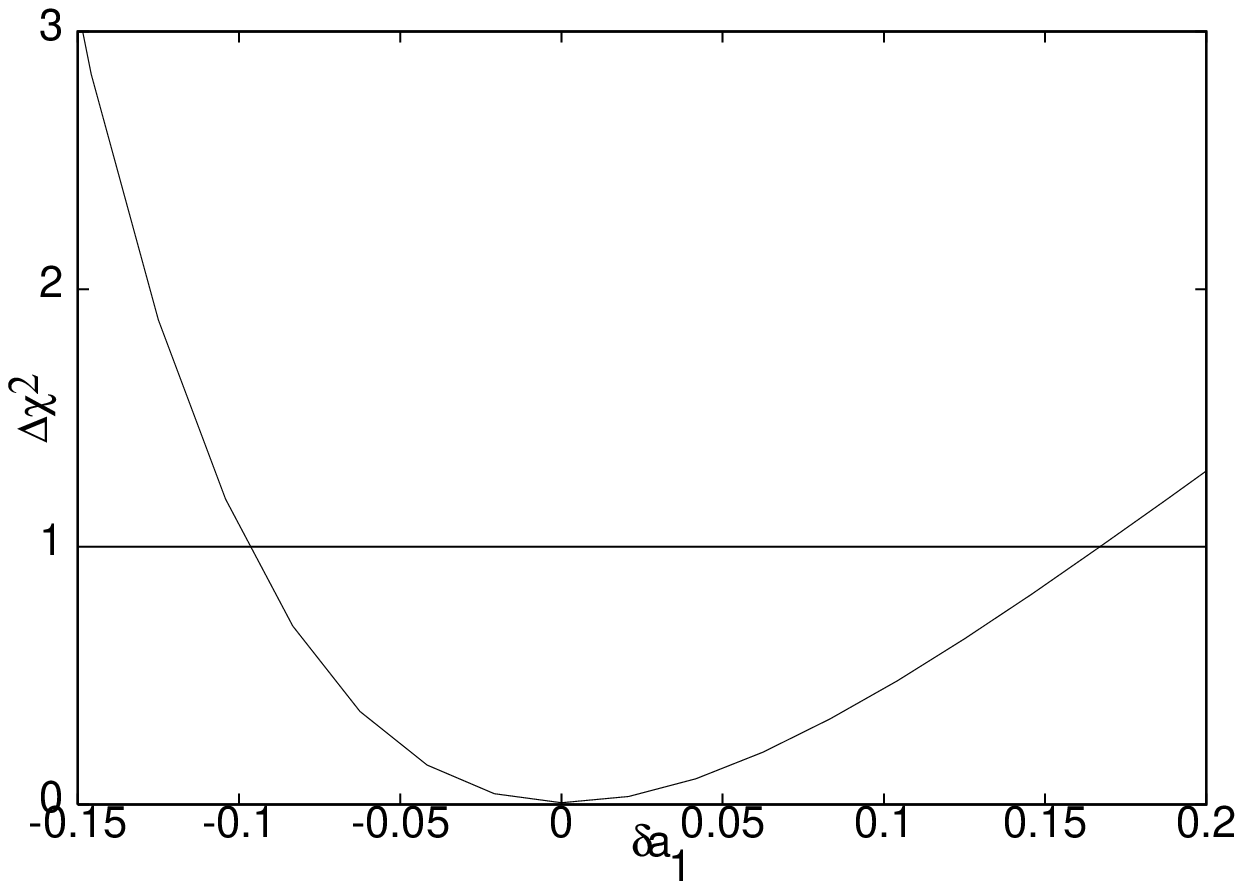,height=5cm,width=7cm}
\end{center}
\caption[dummy]{Plot of $\Delta\chi^2$ against the deviation from the
   best fit value of the gluon parameters $a_0$ and $a_1$ using faked
   data with 20-30\% errors in measurements of $A_1$ for $x<0.1$.}
\label{fg.fake}\end{figure}

We investigated this question quantitatively by generating fake data
at $x<0.1$ from the NLO S set.  The values of $A_1^p$ so generated were
smeared randomly over a 10\% band to simulate noise in the data, and error
bars of 20--30\% were assigned to each such data point. This faked set
is meant to mimic data that could possibly come from a future polarised
HERA experiment.  We redid our fits with this faked data set replacing
all data on $A_1^p$ for $x<0.1$. This data brings down the error bars in
the parameter $a_0$ appearing in $\widetilde g$ by a factor of 4, and
improves the errors in $a_1$ by a factor of 2--3.  The error estimates
obtained with the fake data set are shown in Figure \ref{fg.fake}. The
region of parameter space allowed by this fake data is shown as the grey
patch in Figure \ref{fg.glncr}. Taking data at $x<0.01$ or over a larger
range of $Q^2$, both of which would be feasible at polarised HERA, would
constrain $\widetilde g$ even better \cite{deroeck}. The ability of RHIC
to fix the gluon densities is, of course, well appreciated. However,
if DIS experiments can fix the gluon densities better, then polarised
RHIC can be used as a discovery machine.

\section{Applications}

\subsection{Flavour asymmetry}

Unlike previous fits of parton densities which had built in the constraint
$\widetilde{\bar d}\approx\widetilde{\bar u}$ \cite{prevfit}, we have
allowed for sea quark densities that violate flavour $SU(2)$ symmetry. The
fits show that the data tolerate, and even prefer, strong flavour symmetry
violations. This is easily seen in the first moments of various densities
(Table \ref{tb.mom1}). Since $\Gamma_0$ and $\Gamma_8$ are small, it is
clear that the pattern $|\Gamma_s|\ll|\Gamma_{\bar u}|$ and $\Gamma_{\bar
d}\approx-\Gamma_{\bar u}$ must follow whenever $\Gamma_3$ is large.

It has been suggested \cite{wasym} that $SU(2)$ flavour asymmetry in the
sea be observed through two combinations of cross sections for production
of $W^\pm$ in longitudinally polarised pp scattering---
\begin{equation}
   A_{\sigma,\delta} =
       \frac{\sigma(W^+,\uparrow\uparrow)\pm\sigma(W^-,\uparrow\uparrow)
             -\sigma(W^+,\uparrow\downarrow)\mp\sigma(W^-,\uparrow\downarrow)}{
              \sigma(W^+,\uparrow\uparrow)\pm\sigma(W^-,\uparrow\uparrow)
             +\sigma(W^+,\uparrow\downarrow)\pm\sigma(W^-,\uparrow\downarrow)},
\label{wasym}\end{equation}
where $A_\sigma$ ($A_\delta$) is defined with the upper (lower)
signs.  At LO these asymmetries can be written as the ratio of certain
combinations of polarised and unpolarised parton densities. At LO the
asymmetry $A_{DY}$ for Drell-Yan pairs can also be written in terms
of the parton densities. At zero rapidity, $A_{DY}$ is a function of
$\sqrt\tau=M/\sqrt S$, where $M$ is the mass of the pairs and $\sqrt
S$ is the center of mass energy of the colliding protons.  For $W^\pm$
production at zero rapidity, the parton densities have to be evaluated
at $M=M_W$, and hence $\sqrt S=M_W/\sqrt\tau$.

\begin{figure}[hbtp]
\begin{center} \leavevmode
   \psfig{figure=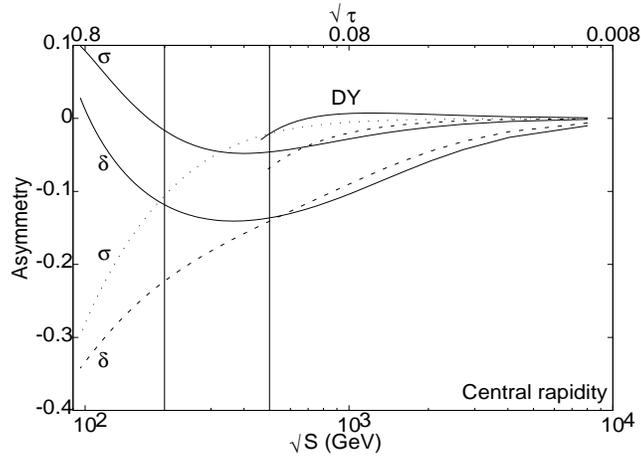,height=6cm,width=8.5cm}
\end{center}
\caption[dummy]{The asymmetries $A_{DY}$ and $A_\sigma$ and $A_\delta$
    (defined in eq.\ \protect\ref{wasym}) computed at LO. The full lines
    are obtained with LO $\overline{\rm S}$ and the dotted lines with LO
    S. The bottom scale is for $A_{\delta,\sigma}$ as a function of
    $\sqrt S$ and that at the top for $A_{DY}$ as a functions of
    $\sqrt\tau$. The vertical band marks out the range $200{\rm\ GeV}\le\sqrt
    S\le500$ GeV.}
\label{fg.wasym}\end{figure}

In Figure \ref{fg.wasym} we have shown these asymmetries at zero rapidity
as a function of $\sqrt S$, or equivalently of $\sqrt\tau$.  It is clear
that at $\sqrt S$ appropriate to RHIC, the iso-triplet spin asymmetry,
$A_\delta$, is best suited to distinguish the LO S densities from LO
$\overline{\rm S}$. Experimentally studying the dependence of $A_\delta$
on $\sqrt S$ over even a limited range below $\sqrt S=500$ GeV would
be very useful. $A_{DY}$ is the least suitable measurement for making
this distinction.

We have already pointed out that the LO fit is unable to decide on the
sign of $\Gamma_g$, and that the NLO fits yield a negative $\Gamma_g$,
although positive values are not ruled out.  Although previous fits have
seen overlapping ranges of allowed $\Gamma_g$, the theoretical bias has
been to take large and positive values of $\Gamma_g$.  This sign can
be easily fixed by various experiments at RHIC or in charm production
measurements at HERA \cite{herapol} or the COMPASS experiment in CERN
\cite{compass}.

We would like to caution that parton densities are renormalisation scheme
dependent (and hence unphysical). They are universally applicable to
all experiments, as long as each experiment is treated in the same
scheme \cite{manohar}. Our determination of these densities are in
the $\overline{\rm MS}$ scheme, and statements about their moments are
therefore also restricted to this scheme. When interpreting the moments of
unphysical parton densities, their scheme dependence must be held in mind.

\subsection{Structure functions and couplings}

It is possible to construct physical quantities out of the unphysical
first moments of the parton densities.  For the first moments of the
structure functions as given in eq.\ (\ref{nloap}), we obtain the values---
\begin{eqnarray}
\nonumber
    &\Gamma_1^p=0.136\pm0.008,\quad&\Gamma_1^n=-0.053\pm0.008\quad({\rm NLO\ }\overline{\rm S}),\\
    &\Gamma_1^p=0.157\pm0.006,\quad&\Gamma_1^n=-0.032\pm0.006\quad({\rm NLO\ S})
\end{eqnarray}
at $Q_0^2=2.56$ GeV${}^2$. These values are within $1.5\sigma$ of each
other and they compare well with values deduced from experiments
\cite{emc,e154,e142}.

We can also use eq.\
(\ref{nloej}) to extract the value of $g_0$. Using as input the above
values of $\Gamma_1^{p,n}$ derived from our fits and the PDG value for
$g_8$, we find
\begin{equation}
    g_0(Q_0^2) = \cases{
            0.13\pm0.20 &(NLO $\overline{\rm S}$),\cr
            0.35\pm0.15 &(NLO S).}
\end{equation}
Since $g_0$ is a physical quantity, it is only to be expected that
our determination of $g_0$ should agree with other analyses, such as
\cite{ab}, even if they use some other scheme to arrive at the same
result.  We will, of course, disagree with them on any scheme
dependent quantity, such as $\Delta\Sigma=\Gamma_u+\Gamma_v+\Gamma_0$.
Our maximally flavour symmetry violating fits give physically reasonable
results.

\subsection{The structure function $g_2$}

Wandzura and Wilczek \cite{WW} have derived a sum rule relating the
twist-2 part of $g_2$ to $g_1$---
\begin{equation}
   g_2^{WW}(x,Q^2) = -g_1(x,Q^2) + \int_x^1 \frac{g_1(y,Q^2)}{y} {\rm d} y.
\label{eq:ww}\end{equation}
An additional twist-2 contribution to $g_2$, suppressed by the ratio of
the quark to the nucleon mass \cite{Song}, is ignored here.  Predictions
for the twist-3 contribution have been made using bag models \cite{bag},
QCD sum rules \cite{sumrule} as well as from non-perturbative lattice
QCD computations \cite{lattice}. Since some computations predict large
twist-3 contributions to moments of $g_2$, it becomes interesting to
check whether the data on $g_2$ allows such contributions.

\begin{figure}[hbtp]
\begin{center} \leavevmode
  \psfig{figure=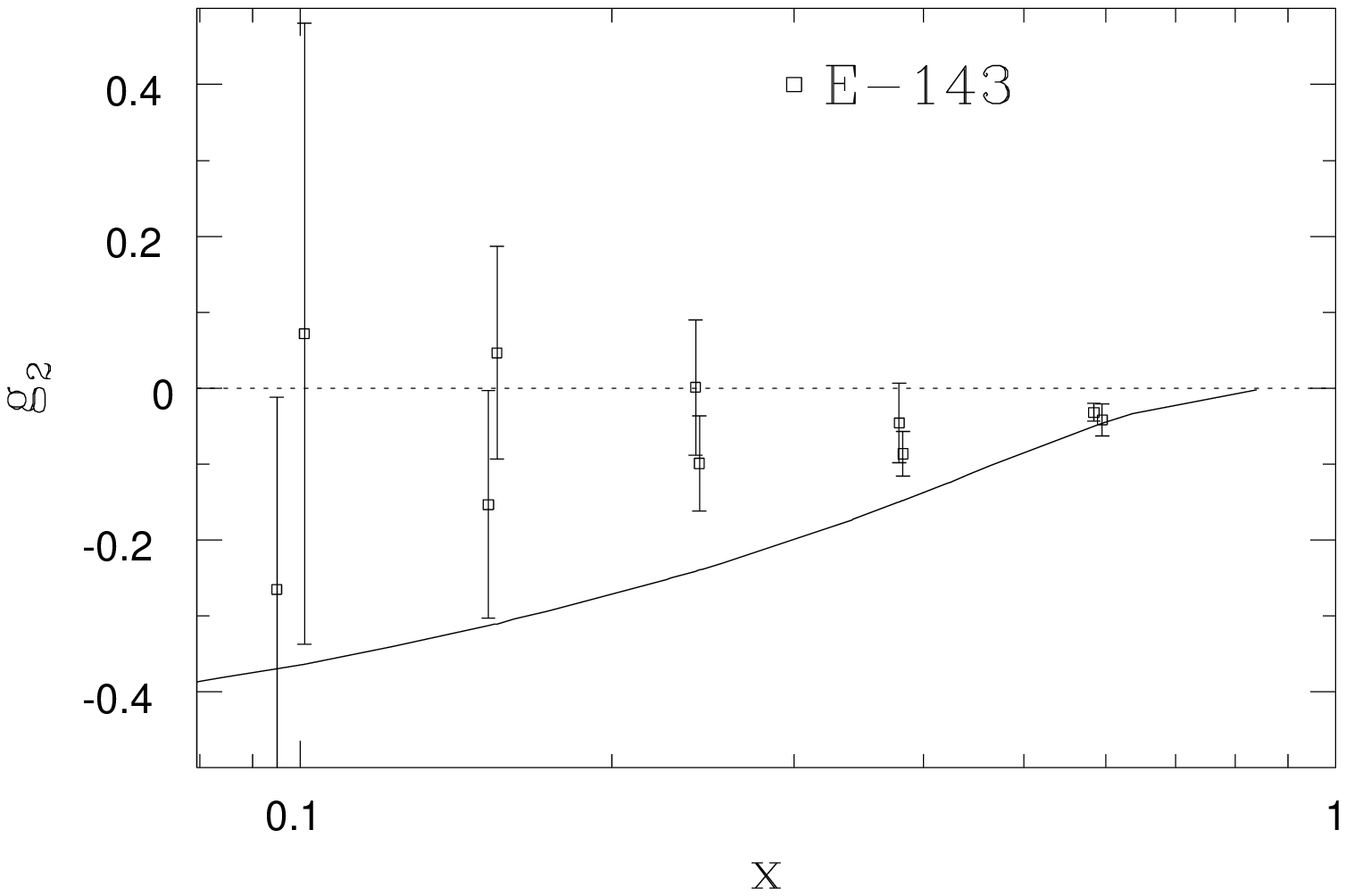,height=4cm,width=4cm}
  \psfig{figure=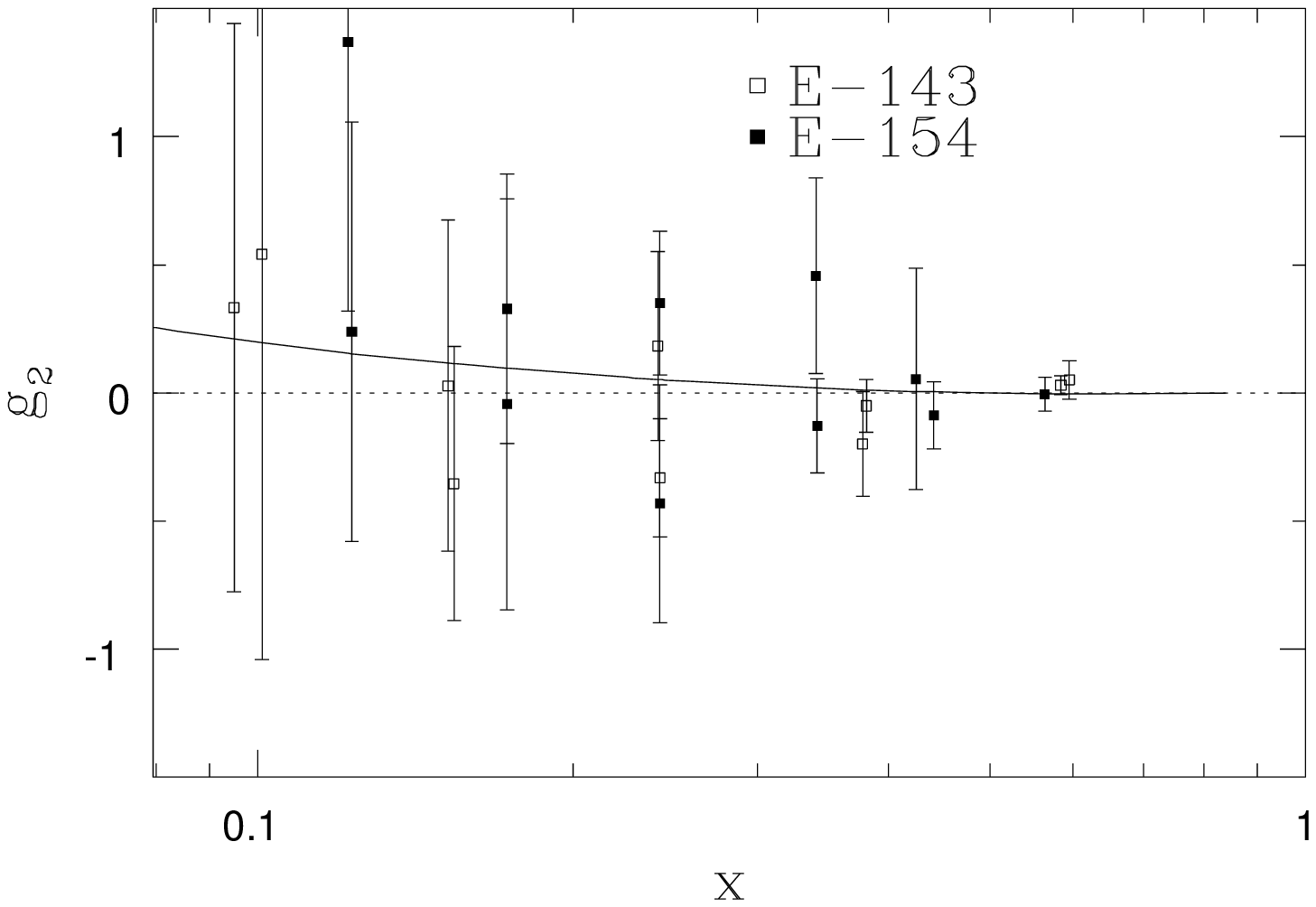,height=4cm,width=4cm}
  \psfig{figure=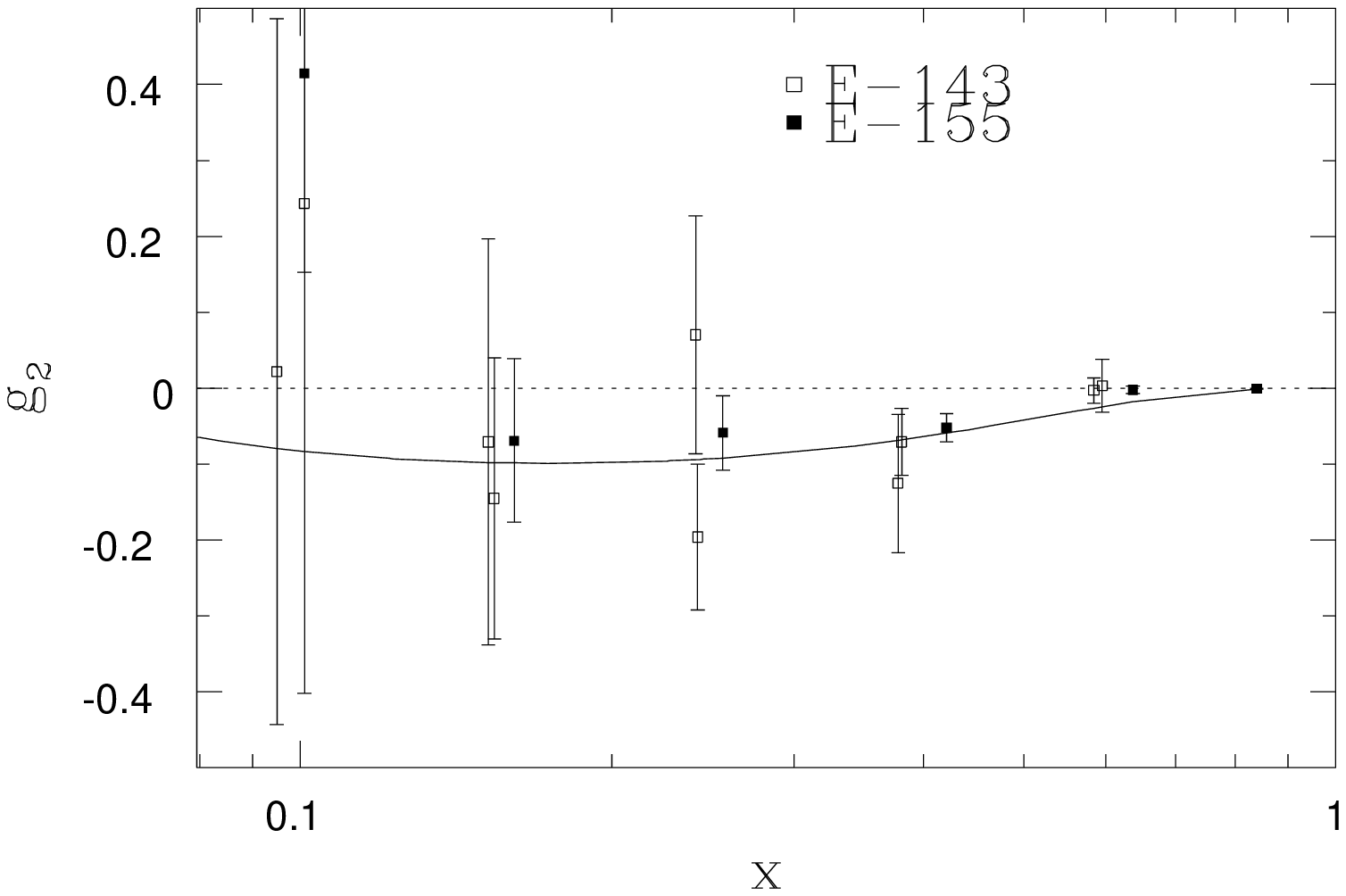,height=4cm,width=4cm}
\end{center}
\caption[dummy]{Data on the structure function $g_2$ compared with the twist-2
   predictions of eq.\ (\ref{eq:ww}) evaluated with our NLO $\overline{\rm S}$
   parametrisation evolved to $Q^2=5$ GeV${}^2$. From left to
   right, the figures are for $g_2^p$, $g_2^n$ and $g_2^d$.}
\label{fg.g2}\end{figure}

Figure \ref{fg.g2} shows our ``prediction'' for the twist-2 part of
$g_2$ and compares it to measurements of this structure function.
Clearly the data is compatible with the NLO twist-2 prediction (and also
with the parton model result, $g_2=0$). Between the prediction and the
data, there is little room for a twist-3 contribution. Statistics have
to be improved vastly in order to study higher-twist effects. In fact,
COMPASS hopes to make this measurement \cite{compass}.

Since the statistical errors are smallest for $g_2^d$, it seems that this
is the best candidate in which to look for twist-3 effects. However the
data quality needs improvement even here. There is considerable scaling
violation in the twist-2 part of $g_2$, but the large errors prevent
any analysis of the $Q^2$-dependence.

\subsection{The valence densities}

\begin{figure}[hbtp]
\begin{center} \leavevmode
   \psfig{figure=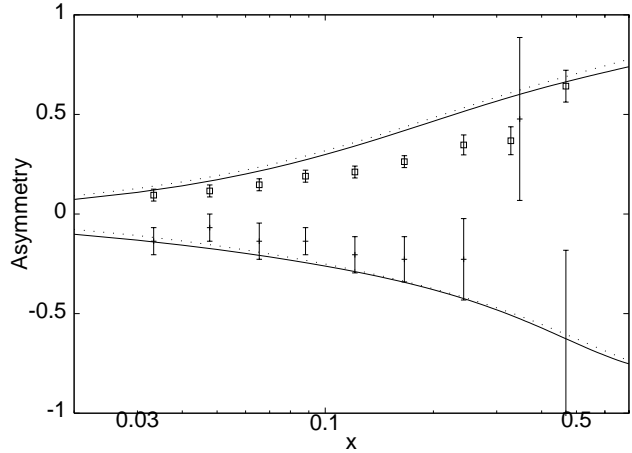,height=6cm,width=8.5cm}
\end{center}
\caption[dummy]{The asymmetries $({\widetilde u}+\widetilde{\bar u})/(u+\bar u)$
   (boxes) and $({\widetilde d}+\widetilde{\bar d})/(d+\bar d)$ (pluses) extracted
   by a parton model analysis of experimental data \cite{semi-h} (the two
   overlapping points at $x=0.35$ have been separated for clarity) compared to
   our NLO fits at $Q^2=5$ GeV${}^2$. NLO $\overline{\rm S}$ is the full line and
   NLO S is the dotted line.}
\label{fg.val}\end{figure}

Recently the HERMES collaboration has used semi-inclusive polarised DIS
data to extract valence and sea quark densities \cite{semi-h}. We have
not used these in our fits because this analysis is performed with parton
model formal\ae. Nevertheless, it is interesting to compare our fits with
these numbers. We display this comparison in Figure \ref{fg.val}. The
rough agreement is heartening, but the small systematic differences
between the fit results and the HERMES extraction of the valence densities
shows the need for a more accurate QCD analysis of the experimental data,
taking into account properly the $Q^2$ dependence through NLO evolution.

\subsection{Axion-matter coupling}

We present an example of the application of polarised proton scattering
to physics beyond the standard model.  The Peccei-Quinn solution to
the strong CP problem postulates a global symmetry whose spontaneous
breakdown generates a (nearly) massless pseudo-Goldstone boson called
the axion \cite{peccei}. There is a variant of the original model which
is still viable \cite{viable}. The axion, $a$, whose decay constant is
$f_a$, couples to fermions, $\psi_f$, of mass $m_f$ by the term
\begin{equation}
   {\cal L}_{int} = -i g_f\,\bar\psi_f\gamma_5\psi_f\,a.
\end{equation}
The coupling $g_f=C_f (m_f/f_a)$. The effective Peccei-Quinn charge, $C_f=
X_f/N$, appears in the coupling instead of the actual charge $X_f$. Here,
$N$ is given by $\sum X_f$. There have been several studies \cite{mele}
of the effective Peccei-Quinn charge of the proton. For three flavours,
the LO expression can be written as
\begin{equation}
   C_{p,n} = \sum_f (C_f-\mu_f)\left[\Gamma_f+\Gamma_{\overline f}\right].
\label{axion}\end{equation}
where $\mu_f=M/m_f$ with $1/M =\sum1/m_f$ \cite{raffelt}.  $C_f$ for
quarks and leptons is highly model dependent. In the so-called KSVZ
\cite{ksvz}, and other hadronic axion models, $C_u=C_d=C_s=0$.  Using
quark mass ratios $m_u/m_d=0.568\pm0.042$ and $m_u/m_s=0.0290\pm0.0043$,
obtained by chiral perturbation theory \cite{leut}, and our LO
$\overline{\rm S}$ fits, we find that
\begin{equation}
   C_p = -0.402\;(2)\qquad{\rm and}\qquad C_n = -0.058\;(4).
\end{equation}
The statistical errors in this coupling are dominated by the errors in
the fits to polarised parton densities. The uncertainty in the quark
masses give smaller contributions to these errors. However, the real
source of uncertainty comes from higher loop corrections. An estimate
of this theoretical uncertainty can be obtained by inserting the NLO
results for various moments into eq.\ (\ref{axion}). This changes $C_p$
by 25\% and $C_n$ by 100\%.

The chiral couplings of neutralinos and charginos in generic
supersymmetric extensions of the standard model also give rise to
effective couplings with matter which depend on the moments of the
parton densities.  Such couplings are often needed in astrophysical
contexts. Unless these couplings are examined to 2-loop order, they
should not be evaluated with the NLO moments.

\section{Conclusions}

We have made global QCD analyses of data on the asymmetry $A_1$
from polarised DIS without making overly restrictive assumptions
about the flavour content of the sea. We have extracted polarised
parton densities which can be used with the CTEQ4 set of unpolarised
densities. Our NLO analyses (in the $\overline{\rm MS}$ scheme) yield
the parameter sets given in Tables \ref{tb.nlo} and \ref{tb.nlo0},
and the LO analyses give the sets displayed in Tables \ref{tb.lo}
and \ref{tb.lo0}.  Strong $SU(2)$ flavour symmetry violation in the sea
is supported by the sets of densities in which the first moment of the
singlet sea quark density is much smaller than that of the triplet sea
quark density.  We also found in the NLO fits that the first moment of
the gluon density is preferentially negative, although positive values
are within 2--3 $\sigma$ of the best fits. The LO fits, to the contrary,
prefer a positive sign for $\Gamma_g(Q_0^2)$, although negative values
are allowed within 2--3 $\sigma$ of the central values.  All physical
quantities obtained from the first moments of our fitted densities have
completely sensible values, as they must have.

We have shown that the polarised HERA option may yield highly accurate
measurements of the polarised gluon densities if measurements of $A_1$
in the range $x\le0.125$ can be performed with errors of about 25\%.
Measurements of the asymmetry in the iso-triplet part of $W^\pm$
production at RHIC are likely to be able to pin down the flavour content
of the sea.  Measurements at future facilities for spin physics thus
nicely complement each other.

We have checked that our parametrisations are roughly consistent with
semi-inclusive DIS data, although a full QCD analysis of this data remains
to be performed.  We have also shown that these parametrisations, when
used to determine the twist-2 part of $g_2$ leave very little room for
a twist-3 part to this structure function. Finally, we have determined
the coupling of hadronic axions to matter--- an input into several
astrophysical constraints on the invisible axion.

In a future publication we plan to make a more detailed study of several
issues, including the proper inclusion of systematic experimental errors
into the analysis and several other technical issues concerning NLO QCD
global fits.

\bigskip\bigskip
We would like to thank Willy van Neerven for several discussions and
clarifications. We also thank the organisers of the 6th Workshop on
High Energy Physics Phenomenology (WHEPP-6), Chennai, January 2000,
where a portion of this work was completed.

\vfil\eject
\begin{table}[hbtp]\begin{center}
  \begin{tabular}{lrrrrr}  \hline
  Experiment & Points
             & LO S & LO $\overline{\rm S}$
             & NLO S & NLO $\overline{\rm S}$ \\
  \hline
  SMC (p)    & 48 & 46.6 & 48.4 & 43.4 & 42.1 \\
  SMC (d)    & 53 & 53.5 & 53.7 & 54.3 & 50.0 \\
  E-143 (p)  & 43 & 48.2 & 47.8 & 48.7 & 50.2 \\
  E-143 (d)  & 43 & 62.1 & 60.3 & 61.0 & 57.0 \\
  HERMES (p) &  9 & 23.8 & 18.8 & 12.6 & 15.1 \\
  HERMES (n) &  4 &  1.7 &  1.6 &  1.5 &  1.7 \\
  E-142 (n)  & 15 & 15.9 & 16.8 & 18.7 & 21.1 \\
  E-154 (n)  &  8 & 22.9 & 16.7 &  3.3 &  2.3 \\
  \hline
  Total      &224 &274.6 &264.0 &243.7 &239.4 \\
  \hline
  \end{tabular}\end{center}
  \caption[dummy]{The contribution to $\chi^2$ from different data sets. The
     parameter sets marked $S$ impose $SU(2)$ flavour symmetry on the
     sea, whereas the sets marked $\overline{\rm S}$ do not.}
\label{tb.chi2}\end{table}

\begin{table}[hbtp]\begin{center}
  \begin{tabular}{cccccc}  \hline
  density        & $a_0$    &  $a_1$  &  $a_2$  & $a_3$ & $a_4$ \\
  \hline
  ${\widetilde V}_u$ &$0.615^{+7}_{-9}$&$-0.32$ ($\pm2$) &$3.689^b$ &12.2 ($\pm2$) &$0.873^b$ \\
  ${\widetilde V}_d$ &$-0.61$ ($\pm2$)&$-0.32^a$&$4.247^b$&2.2 ($\pm1$) &$0.333^b$ \\
  ${\widetilde q}_0$ & 0.009 ($\pm9$) &$-0.2^{+\infty}_{-2}$&$8.041^b$ &8 ($\pm16$) &$1.000^b$ \\
  ${\widetilde q}_3$ &$-0.22^c$&$-0.32^a$  &$8.041^b$ &$0^b$&7 ($\pm5$)\\
  ${\widetilde g}$   &$-1.0^{+3}_{-4}$&$-0.7^{+2}_{-1}$& $4.673^b$ &$-5^{+4}_{-2}$&$1.508^b$ \\
  \hline
  \end{tabular}\end{center}
  \caption[dummy]{The NLO $\overline{\rm S}$ fits for the parameters in eqs.\ (\ref{param},
     \ref{param3}) at $Q_0^2=2.56$ GeV${}^2$. The error estimates shown in the brackets
     apply to the last digit of the estimated value. In case of asymmetric errors, if one
     of the errors is zero it indicates that the parameter is at the limit of positivity.
     The parameters marked (a) are set equal to some other in the same column, (b) are
     fixed to the value taken by the unpolarised densities, and (c) are fixed by the Bjorken
     sum rule.}
\label{tb.nlo}\end{table}

\begin{table}[hbtp]\begin{center}
  \begin{tabular}{cccccc}  \hline
  density        & $a_0$    &  $a_1$  &  $a_2$  & $a_3$ & $a_4$ \\
  \hline
  ${\widetilde V}_u$ &1.74 ($\pm1$)&$-0.149^{+4}_{-3}$&$3.689^b$ &$3.91^{+6}_{-3}$&$0.873^b$ \\
  ${\widetilde V}_d$ &$-0.75^c$&$-0.149^a$&$4.247^b$ &1.6 ($\pm1$)&$0.333^b$ \\
  ${\widetilde q}_0$ &$-0.26$ ($\pm2$)&$-0.08$ ($\pm4$)&$8.041^b$ &6.5 ($\pm7$) &$1.000^b$ \\
  ${\widetilde g}$   &$-0.3^{+1}_{-0}$&$-0.6^{+3}_{-2}$& $4.673^b$ &$-17^{+15}_{-0}$&$1.508^b$ \\
  \hline
  \end{tabular}\end{center}
  \caption[dummy]{The NLO S fits for the parameters in eq.\ (\ref{param}) at
     $Q_0^2=2.56$ GeV${}^2$. Asymmetric errors and superscripts on the numbers have the same
     meaning as in Table \ref{tb.nlo}.}
\label{tb.nlo0}\end{table}

\begin{table}[hbtp]\begin{center}
  \begin{tabular}{cccccc}  \hline
  density        & $a_0$    &  $a_1$  &  $a_2$  & $a_3$ & $a_4$ \\
  \hline
  ${\widetilde V}_u$ &$1.65^{+2}_{-0}$&$-0.159^{+0}_{-6}$&$3.465^b$&$4.3^{+1}_{-0}$&$1.146^b$ \\
  ${\widetilde V}_d$ &$-0.75^{+0}_{-2}$&$-0.159^a$       &$4.003^b$&$2.0^{+1}_{-0}$&$0.622^b$ \\
  ${\widetilde q}_0$ &$-0.21$ ($\pm2$)&$0.01^{+5}_{-4}$&$6.877^b$&$0.6^{+7}_{-6}$& $1.000^b$ \\
  ${\widetilde q}_3$ &$0.81^c$& $-0.159^a$  & $6.877^b$ & $0^b$    & $-3.5^{+4}_{-0}$ \\
  ${\widetilde g}$   &$-0.16^{+5}_{-0}$&$-1.0^{+1}_{-0}$&$3.666^b$&$-15^{+13}_{-0}$&$1.968^b$ \\
  \hline
  \end{tabular}\end{center}
  \caption[dummy]{The LO $\overline{\rm S}$ fits for the parameters in eqs.\ (\ref{param},
     \ref{param3}) at $Q_0^2=2.56$ GeV${}^2$. Asymmetric errors and  superscripts on the
     numbers have the same meaning as in Table \ref{tb.nlo}.}
\label{tb.lo}\end{table}

\begin{table}[hbtp]\begin{center}
  \begin{tabular}{cccccc}  \hline
  density        & $a_0$    &  $a_1$  &  $a_2$  & $a_3$ & $a_4$ \\
  \hline
  ${\widetilde V}_u$ &1.91 ($\pm1$)&$-0.150$ ($\pm3$)&$3.465^b$&$3.44^{+6}_{-7}$ & $1.146^b$ \\
  ${\widetilde V}_d$ &$-1.22^c$&$-0.150^a$& $4.003^b$ &$0.5^{+4}_{-3}$& $0.622^b$ \\
  ${\widetilde q}_0$ &$-0.20$ ($\pm2$)&$0.03^{+5}_{-4}$& $6.877^b$ & $2.2^{+0}_{-5}$ & $1.000^b$ \\
  ${\widetilde g}$   &$-0.16^{+1}_{-0}$&$-1.0^{+2}_{-0}$&$3.666^b$&$-15^{+25}_{-0}$&$1.968^b$\\
  \hline
  \end{tabular}\end{center}
  \caption[dummy]{The LO S fits for the parameters in eq.\ (\ref{param}) at
     $Q_0^2=2.56$ GeV${}^2$. Asymmetric errors and superscripts on the numbers have the same
     meaning as in Table \ref{tb.nlo}.}
\label{tb.lo0}\end{table}

\begin{table}[hbtp]\begin{center}
  \begin{tabular}{ccccc}  \hline
                & LO S & LO $\overline{\rm S}$ & 
                 NLO S & NLO $\overline{\rm S}$ \\
  \hline
  $\Gamma_u$ & 0.875 ($\pm5$)    & 0.829 ($\pm1$)    & 0.909 ($\pm8$)    & 0.85 ($\pm3$) \\
  $\Gamma_d$ & $-0.40$ ($\pm4$)  & $-0.338$ ($\pm6$) & $-0.36$ ($\pm1$)  & $-0.52$ ($\pm3$) \\
  $\Gamma_0$ & $-0.029$ ($\pm3$) & $-0.028$ ($\pm3$) & $-0.058$ ($\pm6$) & $-0.003$ ($\pm3$) \\
  $\Gamma_g$ & -                 & -                 & $-0.2^{+2}_{-6}$  & $-1.6\pm1.0$ \\
  $\Gamma_3$ & -                 & 0.107 ($\pm3$)    & -                 & $-0.10$ ($\pm2$) \\
  $2\Gamma_{\bar u}$ & $-0.0059$ ($\pm6$) & 0.0212 ($\pm9$) & $-0.012$ ($\pm1$) & $-0.024$ ($\pm6$) \\
  $2\Gamma_{\bar d}$ & $-0.0059$ ($\pm6$) & $-0.0323$ ($\pm9$) & $-0.012$ ($\pm1$) & 0.024 ($\pm6$) \\
  $2\Gamma_{\bar s}$ & $-0.0029$ ($\pm3$) & $-0.0027$ ($\pm3$) & $-0.0058$ ($\pm6$) & $-0.0003$ ($\pm3$) \\
  \hline
  \end{tabular}\end{center}
  \caption[dummy]{Moments of various densities at $Q_0^2=2.56$ GeV${}^2$. By our
     initial conditions $\Gamma_8=2\Gamma_0/5$. The numbers in brackets are the
     errors on the last digit of the central value. $\Gamma_g$ is essentially
     undetermined at LO.}
\label{tb.mom1}\end{table}

\end{document}